\documentclass[10pt,final,twocolumn]{IEEEtran}

\ifCLASSINFOpdf
\else
  
\fi

\usepackage{graphicx}
\usepackage{subfigure}
\usepackage{amsmath}
\usepackage{dsfont}
\usepackage{algorithmic}
\usepackage{xspace}
\usepackage[ruled,vlined]{algorithm2e}
\usepackage{color}
\usepackage[font=small]{caption}
\usepackage{mathrsfs}
\usepackage{pdflscape}
\usepackage{array,multirow,graphicx}

\usepackage{times}
\usepackage{psfrag}
\usepackage{amssymb,amsmath}
\usepackage[innercaption]{sidecap}
\usepackage{bm}
\usepackage{url}

\def\Blue#1{\textcolor{black}{#1}}

\begin{document}

\title{Wireless Avionics Intra-Communications: \\ A Survey of Benefits, Challenges, and Solutions}

\author{Pangun Park, Piergiuseppe Di Marco, Junghyo Nah, and Carlo Fischione \thanks{P. Park is with the Department of Radio and Information Communications Engineering, Chungnam National University, Korea (e-mail: pgpark@cnu.ac.kr). P. Di Marco is with the Department of Information Engineering, Computer Science and Mathematics, University of L'Aquila, Italy (e-mail: piergiuseppe.dimarco@univaq.it). J. Nah is with the Department of Electrical Engineering, Chungnam National University, Korea (e-mail: jnah@cnu.ac.kr). C. Fischione is with the Department of Network and Systems Engineering, School of Engineering, KTH Royal Institute of Technology, Sweden (e-mail: carlofi@kth.se).}}


\IEEEcompsoctitleabstractindextext{
\begin{abstract}
In the aeronautics industry, wireless avionics intra-communications have a tremendous potential to improve efficiency and flexibility while reducing the weight, fuel consumption, and maintenance costs over traditional wired avionics systems. This survey starts with an overview of the major benefits and opportunities in the deployment of wireless technologies for critical applications of an aircraft. The current state-of-art is presented in terms of system classifications based on data rate demands and transceiver installation locations. We then discuss major technical challenges in the design and realization of the envisioned aircraft applications. Although wireless avionics intra-communication has aspects and requirements similar to mission-critical applications of industrial automation, it also has specific issues such as complex structures, operations, and safety of the aircraft that make this area of research self-standing and challenging. To support the critical operations of an aircraft, existing wireless standards for mission-critical industrial applications are briefly discussed to investigate the applicability of the current solutions. Specifically, IEEE 802.15.4-based protocols and Bluetooth are discussed for low data rate applications, whereas IEEE 802.11-based standards are considered for high data rate applications. Eventually, we propose fundamental schemes in terms of network architecture, protocol, and resource management to support the critical avionics applications and discuss the research directions in this emerging area.
\end{abstract}
\begin{IEEEkeywords} 
Wireless Avionics Intra-Communications, Industrial Wireless Networks, Real-time Systems, Mission-critical Communications.  
\end{IEEEkeywords}}


\maketitle

\IEEEdisplaynotcompsoctitleabstractindextext
\IEEEpeerreviewmaketitle

\section{Introduction}  
Traditionally, the aircraft control systems rely on expensive wired fieldbus networks to guarantee the flight safety requirements~\cite{Collins06, WAIC_obj}. Wired fieldbus costs include the cable harness design, the labor-intensive cable manufacturing, and the operating and maintenance costs of fibers and connectors~\cite{Furse01, wire1, wire2}. In particular, the critical flight control systems require complex redundant fieldbus channels where the channels are physically and electrically separated from each other to improve fault tolerance~\cite{Yeh01}. For instance, a large commercial transport airplane like Boeing 747 includes roughly 228~km of wire,  which weighs approximately 1,587~kg~\cite{wire1}. Furthermore, recent advanced technologies such as micro sensors and integrated modular avionics architecture need even more bandwidth and more flexible fieldbus topologies~\cite{Prisaznuk08,IMA}. This shows that we need a fundamentally new network infrastructure to reduce installation and maintenance costs and environmental impacts while meeting the safety requirements for next generation avionics systems~\cite{nextgen}. 

As one of the novel solutions, the aviation industry strives to use wireless technologies in both current aircraft upgrades and new aircraft design. The reason is that Wireless Sensor Networks (WSNs) and Wireless Local Area Network (WLAN) have been successfully used for realizing industrial monitoring and control systems~\cite{Kumar14, Watteyne16}. Wireless Avionics Intra-Communication (WAIC) systems can significantly improve the operational efficiency and flexibility over current wired systems on the aircraft~\cite{WAIC_obj,Elgezabal10,iwlan1}. WAIC is restricted to applications related to secure, reliable, and effective aircraft operations such as structural health monitoring, sensing, control, voice, video, and fieldbus communications, as defined by the International Civil Aviation Organization (ICAO)\footnote{The International Civil Aviation Organization develops standards and manuals to ensure safety and growth for civil aviation.}~\cite{WAIC_obj}. However, WAIC systems are not designed to provide communications for in-flight entertainment or with consumer devices carried onboard the aircraft. Furthermore, WAIC systems do not support any aircraft-to-ground, aircraft-to-aircraft, and aircraft-to-satellite communications. In fact, the wireless avionics technology has its origins onboard the international space station~\cite{invocon1}. NASA obtains the mechanical motion of the station while coupled to the space shuttle using wireless communication from external sensors~\cite{invocon2}. In~2008, Gulfstream aerospace corp. has successfully completed the feasibility test of the concept called ``Fly-by-Wireless" systems~\cite{Gomez10, Gulfstream}. 


\Blue{Because WAIC systems necessarily operate worldwide and cross national borders due to the international nature of air travel, different international organizations, government agencies, and independent aviation industry cooperate in addressing several technical issues.} The Aerospace Vehicle Systems Institute (AVSI)\footnote{The Aerospace Vehicle Systems Institute, a cooperative of companies, academia and government agencies, has played a major role in the overall development, classification and standardization of WAIC since 2008.} has extensively investigated the total spectrum demand to support various WAIC applications and has analyzed the compatibility and interference issue with existing aeronautical radionavigation service below $15.7$~GHz~\cite{WAIC_freq}.  In 2015, the International Telecommunication Union - Radiocommunication (ITU-R) finally announced the $4.2-4.4$~GHz frequency as the primary spectrum band for WAIC systems at the world radio conference~\cite{WRC15}. Hence, Federal Aviation Administration, European Union Aviation Safety Agency, and ICAO  rely on the assigned frequency to develop and deploy WAIC systems. 



\begin{figure}[t]
  \centering
  \includegraphics[width = 0.98 \columnwidth]{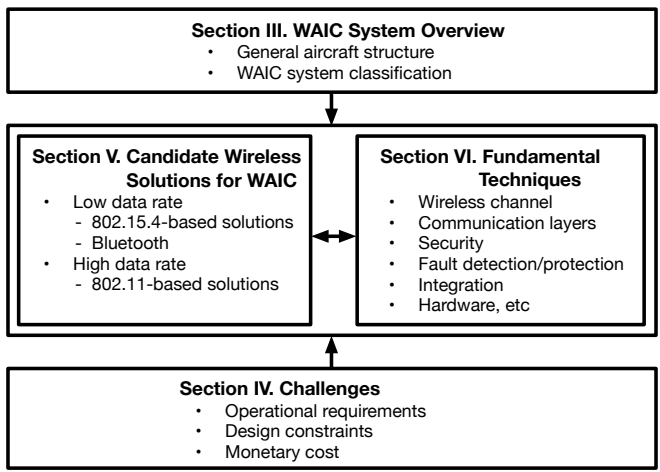}  %
   \caption{\Blue{Overall organization of the paper: Representation of the different WAIC systems described in Section~\ref{sec:waic}, the major technical challenges analyzed in Section~\ref{sec:cha}, and the most suitable standards and techniques illustrated in Sections~\ref{sec:existing} and~\ref{sec:tech}, respectively.}} \label{fig:org}
\end{figure}


\Blue{The goal of this survey is to provide an overview of the WAIC framework, main challenges, and promising fundamental techniques to design, analysis, and deployment of WAIC systems.} Major contributions of the paper are as follows. 
\begin{itemize}
\item We introduce the major benefits of WAIC system over traditional fieldbus infrastructure for avionics systems in Section~\ref{sec:benefit}. 

\item We provide a general overview of the current state of WAIC frameworks in Section~\ref{sec:waic}. 

\item We discuss the major technical challenges in WAIC systems for realizing the envisioned critical avionics applications in Section~\ref{sec:cha}.  

\item We describe closely related industrial wireless standards as possible candidates for WAIC systems in Section~\ref{sec:existing}. 

\item Finally, we discuss the fundamental technical schemes to resolve the technical challenges and the still-open research issues in Section~\ref{sec:tech}. 
\end{itemize}

\Blue{Fig.~\ref{fig:org} summarizes the major relationships between complex systems of Section~\ref{sec:waic}, technical challenges of Section~\ref{sec:cha}, and the most suitable standards and techniques illustrated in Sections~\ref{sec:existing} and~\ref{sec:tech}, respectively.} To the best of our knowledge, this is the first study that thoroughly discusses the main benefits, technical challenges, and design principles to develop and deploy WAIC systems.



\section{Major Benefit}\label{sec:benefit}
By replacing cable, WAIC offers competitive advantages such as cost saving in terms of fuel consumption, installation, and maintenance, and improvement of reliability, flexibility, and scalability~\cite{Gomez10, Bur10}. In the following, we elaborate more in the details about the major advantages of WAIC systems.

\subsection{Cost Saving} 
The installation, operation, and maintenance costs of cabling result in a considerable pressure to the aircraft manufacturer and the airline operators~\cite{Collins06, Alena07}. Due to the expense of materials and labor, the cost of installing sensors and fieldbuses is likely not decreasing in the future.

\begin{figure*}[t]
  \centering
  \includegraphics[width = 1.95 \columnwidth]{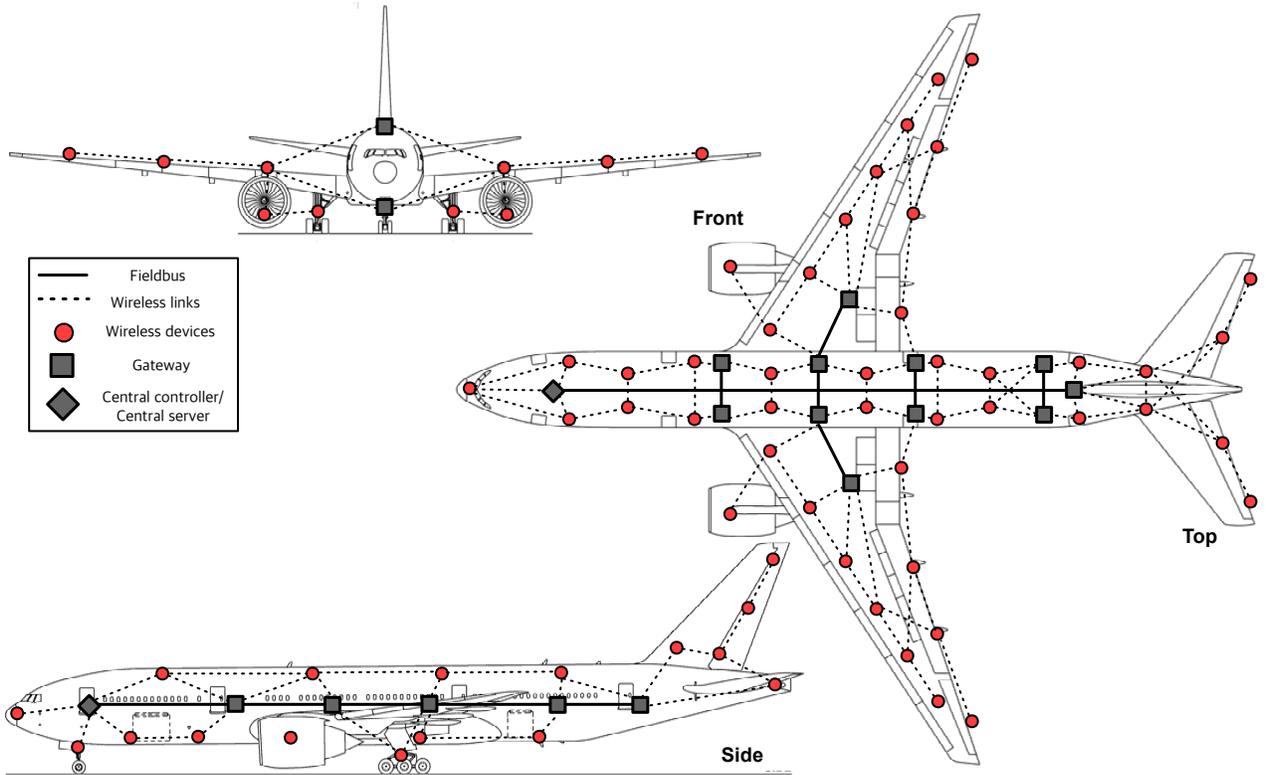}  %
   \caption{Proposed conceptual wireless avionics intra-communications composed of a large number of wireless devices and wired backbone for a typical passenger aircraft.}
   \label{fig:waic}
\end{figure*}

One of the critical factors to develop a new aircraft is the wiring harness design~\cite{Furse01}. In fact, the typical cost of the sensor installation is usually several times the actual cost of the sensor module itself. The developers must indicate and determine the complicated paths onboard the aircraft for kilometers of wire. The redundant wiring must ensure separate routing paths so that redundant circuits are not affected by a single point failure~\cite{Gomez10}. As the safety demands become more strict, the cable weight is increasing due to the higher redundant resource requirements. Furthermore, diagnosing issues within an advanced high-speed fieldbus network require highly skilled technicians~\cite{Schuster08}.

For instance, the total cabling weights of A320/B737 and A350/B787 are around 15\% and 20\% of the total weight of aircraft, respectively~\cite{Furse01,Elgezabal10}. In each aircraft, the related costs including manufacturing and installation are approximately \$2,200 per kilogram, resulting in a cost varying from \$14~million for A320/B737 to \$50~million for A350/B787~\cite{Elgezabal10}. Cabling cost further increases for new generation aircraft such as the A380 because of the longer cables. In A380, the extremely long cable of around 500~km is the main reason of the cost overruns and production delays, estimated at \$2~billion.

Wireless networks are an appealing technology since they can substantially decrease the time and cost of cable harness design and installation, and eventually life-cycle costs of aircraft. Since wireless links provide redundant connectivity without specific redundant cables, it is a cost effective solution for various types of aircraft. The price of wireless sensors and network devices is rapidly coming down following Moore's Law~\cite{Kumar14}. Furthermore, it is estimated that wireless communication can substantially decrease the fuel consumption by 12\% as a result of the weight reduction~\cite{Elgezabal10}.

\subsection{Availability}  
Avionics systems must provide high availability and determinism in the flight control systems to avoid catastrophic consequences such as injury, explosions, and human losses~\cite{Buckwalter08},~\cite{Schmitt98}. Wiring is a major source of maintenance and failure cost since it can affect the immunity of the interconnected system by inducing more than 50\% of electromagnetic interference within the aircraft~\cite{Collins06, Furse01}. Such defects particularly occur at interface components such as connectors, pins, and sockets and are extremely difficult to resolve and repair~\cite{Collins06}. A number of aircraft, including TWA 800 and Swissair 111, have been lost in the past due to wiring failures~\cite{Panitz10}. Due to wiring discrepancies, the U.S. Navy has $78$ non-mission capable aircraft, about 27,365 flight hours between aborts in 2005~\cite{Collins06}. A wireless system can improve the availability by substantially reducing electrical interfaces of the system. 

Redundant components of cable harnesses are the main techniques to achieve a fault-tolerant aircraft design~\cite{Yeh01,Buckwalter08}. However, duplicated cables using the identical technology is generally vulnerable to common failures such as lightning strike or fire. The wireless system as a backup to a typical fieldbus system provides redundancy through diverse techniques. Hence, wireless networks offer the cost effective built-in redundancy instead of complicated wired connections to the avionics systems.

\subsection{Flexibility}
WAIC can support some new applications such as monitoring rotating unit, enabling mobile workers for maintenance, and integration of non-traditional signals including voice, image, and video. In particular, WAIC can collect information from where it was technically infeasible. For instance, one interesting application is the bearing monitoring of engine rotators, that cannot be performed with wiring harnesses. 

Due to significant pressure changes, the fuselage health is one of the most critical factors to guarantee the long average lifespan of civil aircraft, which typically exceeds 25 years~\cite{Elgezabal10}. Throughout the life-cycle of aircraft, WAIC could considerably reduce the complexity and cost to mount new sensors, and allow easier system modifications~\cite{Elgezabal10}. For example, on-board sensors monitor lightning or other environmental damage during the flight operation. By using WAICs, it is not necessary to route the fieldbus cable to the dedicated controller from each sensor and actuator~\cite{Bur10}. Furthermore, it supports flexible cabin configurations for more customized subsystems.



\begin{table}[t]
\centering
\begin{tabular}{ |c|c|c| } 
 \hline
 Level & Failure condition & Failure rate \\  \hline 
A & Catastrophic & $10^{-9}$/h \\  \hline 
B & Hazardous & $10^{-7}$/h \\  \hline 
C & Major & $10^{-5}$/h \\  \hline 
D & Minor & $10^{-3}$/h \\  \hline 
E & No effect & n/a\\ \hline 
\end{tabular}
\caption{Development Assurance Level (DAL) indicates the level of safety-critical software function of an aircraft based on the safety assessment process in DO-178B~\cite{DO178B}. DO-178B is a guideline for the safety-critical software certification used in airborne systems.} \label{tab:DO178}
\end{table}

\section{WAIC system overview}\label{sec:waic}  
\Blue{Fig.~\ref{fig:waic} presents the conceptual WAIC network consisting of a large number of wireless devices and wired backbone. Wireless devices include various sensors, actuators, and relay nodes of an aircraft. WAIC must support heterogeneous applications from monitoring systems to flight control systems, such as structural health monitoring, sensing, control, voice, video, and fieldbus communications. In contrast to the typical monitoring application of industrial WSNs, the heterogeneous applications include both low data rate application such as temperature measurements in cabins and high data rate applications such as video surveillance~\cite{Kumar14,Park18}. The heterogeneous software used in airborne systems has various criticality levels and operational requirements. The software level, called the Development Assurance Level (DAL), classifies various safety-critical software into five levels based on the impacts of system failure~\cite{DO178B}. Table~\ref{tab:DO178} shows five classes of the safety-critical software in airborne systems.}

This section provides the overview of the WAIC framework. We first briefly introduce a reference aircraft model. We then present a classification of WAIC systems based on the data rate demand and transceiver installation location.


\begin{figure}[t]
  \centering
  \includegraphics[width = 0.99\columnwidth]{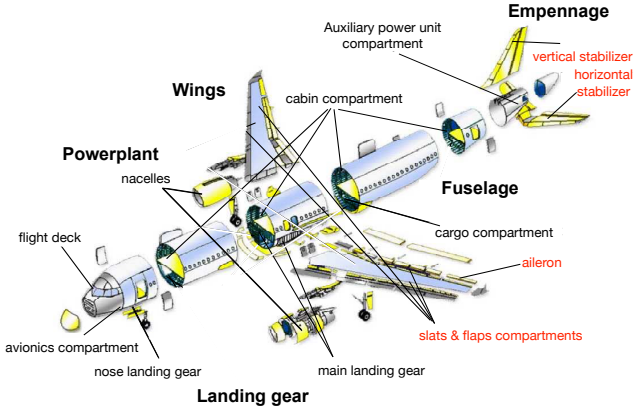}
   \caption{Main components of a typical passenger aircraft with multiple compartments~\cite{WAIC_obj}. Most aircraft structures consist of a fuselage, an empennage, a powerplant, wings, and landing gears~\cite{Gud14_struct}. We emphasize the main actuators located around wings and empennages for flight controls in red.} \label{fig:RefAir}
\end{figure}

\subsection{General Aircraft Structure} 
Fig.~\ref{fig:RefAir} illustrates the main structural parts of a typical passenger aircraft with the major components and various compartments~\cite{WAIC_obj}. Most aircraft structures consist of a fuselage, an empennage, a powerplant, wings, and landing gears~\cite{Gud14_struct}. The fuselage is the centrally located main part including the cockpit, cabin, and cargo compartments. The empennage includes the vertical and horizontal stabilizers at the entire tail part. These movable surfaces are used to control the horizontal rotation and the vertical rotation of the aircraft. The wings are the primary lifting surfaces attached to each side of the fuselage including several critical actuators such as ailerons, slats, and flaps to support the flight control systems. The powerplant includes all engine components, the propeller, and electrical system such as nacelles and auxiliary power unit compartment, as shown in Fig.~\ref{fig:RefAir}. The landing gear consists of wheels and struts to support the mobility on the ground. We remark that critical actuators of flight controls are mainly located in wings and tail section of aircraft, while the control-related sensors such as speed sensors are installed around nacelles and fuselage~\cite{Schmitt98,Gud14_struct}. In Fig.~\ref{fig:RefAir}, we emphasize the locations of main actuators for flight controls in red. In addition, structural health monitoring sensors are commonly installed on the main structure of aircraft including wings and fuselage~\cite{Wilson14}. Hence, wireless transceivers of the WAIC system are mounted at various locations both inside and outside the airframe.

The commercial aviation industry considers the passenger aircraft of Fig.~\ref{fig:RefAir} as the main target of WAIC~\cite{WAIC_obj}. A typical passenger aircraft between 150 to 220 seats is considered as the reference aircraft to analyze the requirements~\cite{WAIC_obj,WAIC_spc}. The overall length of this typical passenger aircraft is between $31.5$ and $44.5$~m. Hence, the maximum range between transmitter and receiver is about $50$~m. Some recent unmanned aircraft and military aircraft do not have cabin compartment and nacelles in contrast to the typical structure of the passenger aircraft~\cite{Gud14_struct}. However, they still have similar major components of Fig.~\ref{fig:RefAir}. Hence, the reference model is the most general structure of the typical aircraft since it sufficiently represents other aircraft types with minor modifications.

\begin{figure}[t]
  \centering
  \includegraphics[width = 0.9 \columnwidth]{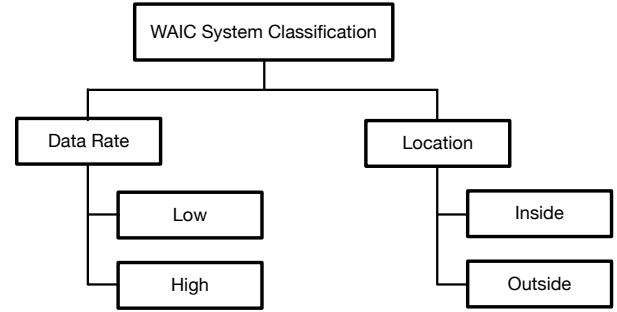} 
   \caption{\Blue{WAIC system classification based on the two features, namely, data rate requirement (high and low) and transceiver location (inside and outside the airframe)~\cite{WAIC_spc}.}} \label{fig:class}
\end{figure}

\begin{table*}[t]
\centering
\begin{tabular}{ |c||c|c|c|c| } 
 \hline
 {} & Low data rate inside & Low data rate outside & High data rate inside  & High data rate outside \\ 
 \hline
Aggregate average data rate of  & \multirow{2}{*}{394} & \multirow{2}{*}{856} & \multirow{2}{*}{18,385} & \multirow{2}{*}{12,300} \\ 
network (kbit/s) &{} & {} & {} & {} \\ 
 \hline
Range of average & \multirow{2}{*}{0.01-0.8} & \multirow{2}{*}{0.02-8} & \multirow{2}{*}{12.5-1,600} & \multirow{2}{*}{45-1,000} \\ 
data rate per link (kbit/s) &{} & {} & {} & {} \\ 
 \hline
Peak data rate per link (kbit/s) & 1 & 8 & 4,800 & 1000 \\ 
 \hline
Number of nodes & 4,150 & 400 & 125 & 65 \\ 
 \hline
Installation domain & inside  & outside & inside & outside \\ 
 \hline
Maximum distance & \multirow{2}{*}{15} & \multirow{2}{*}{15} & \multirow{2}{*}{15} & \multirow{2}{*}{15} \\ 
between TX and RX (meter) & {} & {} & {} & {} \\ 
 \hline
Typical channel & NLOS & LOS/NLOS & LOS/NLOS & LOS \\ 
 \hline
\multirow{4}{*}{Application} & sensing and control  & sensing and control  & sensing and communication & sensing and control \\ 
{} & (cabin temperatures, & (temperature, pressure,   & (engine,  & (structure, \\ 
{} & pressure control, & structural stress, & avionics data bus, & vibration control, \\ 
{} & smoke, door)   & landing gear) & voice/video/image)  & voice/video/image) \\ 
\hline
Most dominant DAL levels & C/D & A/B & B/C & B/C \\ 
\hline
Spectrum requirements  & \multicolumn{2}{c|}{\multirow{2}{*}{35}} & \multicolumn{2}{c|}{\multirow{2}{*}{53}} \\ 
per aircraft (MHz) & \multicolumn{2}{c|}{{}} & \multicolumn{2}{c|}{{}} \\ 
 \hline
Maximum transmit power (mW) & \multicolumn{2}{c|}{10}  & \multicolumn{2}{c|}{50}  \\ 
 \hline
Receiver sensitivity (dBm) & \multicolumn{2}{c|}{-91}  & \multicolumn{2}{c|}{-77}  \\ 
 \hline
\end{tabular}
\caption{Technical characteristics of four classes for WAIC systems~\cite{WAIC_spc}.} \label{tab:class}
\end{table*}



\subsection{WAIC System Classification}\label{sec:classification} 
Current architecture of relatively new aircraft such as the A350 and A380 mainly consists of two different networks, namely, high data rate and low data rate fieldbuses~\cite{Buckwalter08,Ricker17}. A high data rate fildbus such as Avionics Full Duplex Switched Ethernet (AFDX)~\cite{A664} is used as a backbone network in order to connect the avionics subsystems of aircraft. In each avionics subsystem, a low rate data fieldbus such as ARINC429~\cite{A429} and MIL-STD-1553~\cite{M1553} directly connects sensors and actuators.

By considering these observations and signal attenuation, AVSI categorizes WAIC systems based on two features, namely, data rate requirement (high and low) and transceiver location (inside and outside the airframe)~\cite{WAIC_spc}. Fig.~\ref{fig:class} shows four classes, namely, ``Low data rate Inside (LI)", ``Low data rate Outside (LO)", ``High data rate Inside (HI)" and ``High data rate Outside (HO)". Table~\ref{tab:class} lists the main characteristics and requirements of four classes including further attributes associated with each individual class. The classification threshold between low and high data rate applications is $10$~kbit/s~\cite{WAIC_spc}. Most low data rate nodes measure scalar data such as strain, pressure, temperature, and humidity. These devices are usually resource-constrained on computation, storage, and energy. High data rate nodes transmit high resolution measurements such as engine status, image, and video. 

The classification of the installation location depends on the Radio Frequency (RF) attenuation of the airframe material. The nodes are regarded ``Inside" only when surround material gives significant RF attenuation like metal. Depending on a particular operation, some applications may be categorized differently. For instance, when the gear is extended, landing gear sensors will be changed to outside class from inside one. As shown in Table~\ref{tab:class}, most transmissions of general aircraft are inside the aircraft structure (e.g. fuselage, wings). However, some critical sensors and actuators operate outside at least for some time. 

We provide detailed descriptions of each class in the following sections. 
\subsubsection{LI Class}
The LI class is characterized by two main attributes, namely, low data rate ($<$10~kbit/s) and transceiver installation inside metal-like enclosures. This applications include wireless sensing and control signals of slowly varying physical processes, such as smoke sensors, door position sensors, and pressure control. Most LI applications belong to level C and D except some critical smoke and fire detection sensors of Table~\ref{tab:DO178}. 

Most of the LI nodes are active throughout all flight phases including the ground operation. The expected data rates are low due to low sampling rate. For instance, the sampling rate of cabin temperature sensor is around 1 sample per second or less. The number of LI links is around $4150$ where the average data rate per link ranges between 10 bit/s and 800 bit/s, for a typical passenger aircraft~\cite{WAIC_spc}. Since most LI nodes are installed in hidden locations, Non-Line-of-Sight (NLOS) propagation channel is dominant. Hence, WAIC transceivers of different compartments may operate at  the same wireless channel if the metallic or conductive composite wall gives significant signal attenuation. 

\subsubsection{LO Class}  
The LO category includes low data rate monitoring applications such as wheel position and speed sensors for control. Although the average data rate per link is below $10$~kbit/s, the sampling rate can be considerably different dependent on applications. The monitoring application such as door position has low sampling rate around several seconds to a minute whereas control applications such as wheel speed sensing need the anti-skid control at $2.5$~ms~\cite{WAIC_obj}. Most LO applications fall into assurance level A and B of Table~\ref{tab:DO178} since they include safety-critical sensing and actuators such as wheel speed, structural stress, ailerons, slats, and flaps. The number of LO links is around 400 where the average data rate per link ranges between $20$~bit/s and $8$~kbit/s~\cite{WAIC_spc}. 

The LO applications do not gain the benefits of the signal attenuation since they operate outside the airframe in most cases. In Fig.~\ref{fig:waic}, a significant number of LO transceivers are installed on exposed areas of the wing, tail, landing gear and wheel wells~\cite{Schmitt98}. In particular, when the flaps, ailerons and spoilers are activated, many critical sensor and actuator transceivers of the flight control system are exposed at the wings and tail, as illustrated in Figs.~\ref{fig:waic} and~\ref{fig:RefAir}. 

\subsubsection{HI Class} 
The flight deck communication, image, video, engine sensors, and avionics fieldbus belong to the HI class. The source traffic mainly consists of regular periodic traffic such as high resolution engine sensors and irregular traffic bursts such as voice/image/video on-demand services. Engine sensors are located within engine nacelles, while most voice/image/video nodes are located in different compartments, such as the cabin, bays, and flight deck, as shown in Figs.~\ref{fig:waic} and~\ref{fig:RefAir}. Engine prognostic sensors are used to monitor various engine parameters for post-flight analysis on the ground. They are not intended for critical flight control purposes. The expected data rate demands of voice/image/video are in tens, hundreds, and thousands of kbit/s, respectively. Furthermore, these HI applications typically require low latency ($<$ 0.5 s) and low jitter ($<$ 50~ms) to maintain quality of on-demand services. 

Most HI application belong to assurance level B and C of Table~\ref{tab:DO178}. The number of HI links is around 125 where the expected data rate per link ranges between $12.5$~kbit/s and $1.6$~Mbit/s, for a typical passenger aircraft. NLOS paths are the dominant wireless conditions except some specific cases of the Line-of-Sight (LOS) paths within the cabin. 

\subsubsection{HO Class}
The HO category includes structural health monitoring and active vibration control sensors. These sensors require high data rate since they commonly require a high sampling rate with high resolution measurements. For the rotorcraft, flight deck voice and video systems are categorized as outside due to its physical layout in contrast to the typical passenger aircraft~\cite{Gud14_struct}. 

Most HO applications fall into assurance level B and C of Table~\ref{tab:DO178}. The number of HO links is around 65 where the average data rate per link is between $45$~kbit/s and $1$~Mbit/s. Since LOS paths are the dominant wireless conditions in the HO category, WAIC transceivers outside the airframe could cause mutual interference.

 

\begin{figure}[t]
  \centering
  \includegraphics[width = 0.9 \columnwidth]{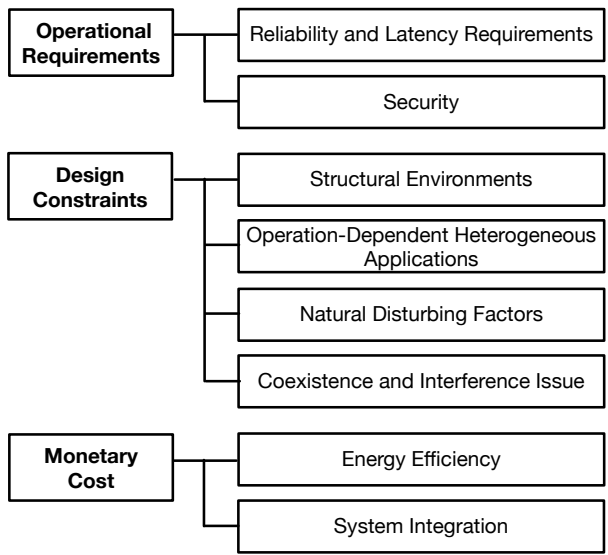} 
   \caption{\Blue{Three primary challenges, namely, operational requirements, design constraints, and monetary cost and their detailed technical problems to design, analysis, and deployment of WAIC systems.}} \label{fig:challenge}
\end{figure}

\section{Challenges}  \label{sec:cha} 
\Blue{As a general guideline, WAIC needs to be cost-efficient while providing comparable real-time and security performance to the current fieldbus technology of aircraft~\cite{Buckwalter08}. In this section, we classify the primary challenges into three categories, namely, operational requirements, design constraints, and monetary cost, as shown in Fig.~\ref{fig:challenge}. The operational requirement is the heart of the essential components to support the real-time and security-critical flight control systems.  Furthermore, WAIC systems are severely affected by design constraints such as the complex structure and the natural disturbance and monetary cost constraints such as system integration and maintenance.}



\subsection{Reliability and Latency Requirements}  
Like all mission-critical systems, WAIC systems have stringent performance demands to ensure the safe operation of an aircraft. Fight control systems require the high reliability and bounded latency of the network to estimate and control the state of the aircraft~\cite{Schmitt98,Buckwalter08}. However, the signal strength of avionics environments is severely affected by noise generated by the avionics, multi-path from the walls, and interference from other aircraft~\cite{WAIC_freq_all}. One of the worst situations is that the pilot of the cockpit does not have a consistent view of the real state of the aircraft due to uncertain network performance. 

The real-time data must be delivered within a relatively short deadline for most control systems of aircraft. To guarantee the stability of closed-loop control systems, both controller and actuator must receive the time-critical sensing data and feedback control signal, respectively, in a timely manner. The WAIC network generally includes a large number of sensing and actuating links, as shown in Table~\ref{tab:class}. In particular, flight control systems consist of a large number of uplinks from sensors to measure speed and altitude with few downlinks to actuators such as ailerons, slats, and flaps~\cite{WAIC_obj}. Uplink reliability is crucial to ensure successful packet transmission such as sensor measurements from wireless node to backbone networks. In addition, downlink reliability must be guaranteed to deliver control or query packets from the backbone to the wireless nodes. In fact, the downlink to actuators turns out to be more important than the one of the uplink in most safety-critical control systems~\cite{Park15}. Hence, a network protocol should consider the reliability in both directions~\cite{Gungor08, Tezcan07, Zhang07}. While the retransmission of data packet improve the reliability at the cost of the delay, outdated packets are generally not useful for control applications~\cite{Park18}. In fact, the reliable and timely requirements of WAIC are more challenging than the ones of the traditional monitoring and open-loop control applications using industrial wireless networks~\cite{Kumar14,Akerberg11}. 



\subsection{Security}
\Blue{Due to its broadcast nature, wireless networks are inherently subject to security threats. In WAIC systems, attacks can come both from omboard (passengers or other entities in the cabin), and from outside the aircraft (ground or other aircrafts). Differently from other WSN systems, data authentication and integrity are essential aspects to guarantee the safety-critical operations of the aircraft.}

\Blue{Security attacks have been analyzed based on an adversary model in the context of wireless-enabled avionics~\cite{Sampigethaya09}. The most relevant attacks for WAIC are summarized in the following.}

\begin{itemize} 
\item \Blue{Jamming attack: Assets of an aircraft can be made unavailable and operation can be disrupted by jamming the wireless medium. Although interference mitigation technique reduces the impact of a jamming attack, prevention of this attack is not entirely possible and the focus is instead on early detection of such attacks.}

\item \Blue{Man-in-the-middle attack: To threaten aircraft safety or airline business, an adversary may attempt to corrupt, insert, or delete assets in the communication path among assets. An example is the manipulation of health diagnostics to prevent fault detection. In these cases, the attach is prevented by limiting physical access to assets by design, however, it needs to be combined with appropriate procedures for device provisioning and commissioning, so that unauthorized access is prohibited.} 

\item \Blue{False alarm: Misleading alarms may not be a threat to aircraft security, but can cause serious economic damages. An adversary can attempt to modify the configuration report to create misalignments between current and intended configurations. Data and configurations need to be protected against replay attacks, so that information cannot be spoofed and reused to alter the configuration.}
\end{itemize} 

\subsection{Structural Environment}
A WAIC system has variable RF propagation characteristics due to the harsh and complex structures of the aircraft with various materials. The basic aircraft structure can be made of either metal or composite materials, or some mixture of the two~\cite{Gud14_mat, air_material}. Graphite epoxy has become a common option for contemporary commercial aircraft, though stainless steel or titanium are still used to endure the higher stress or heat for military aircraft. Note that most WAIC transceivers inside the airframe are installed in hidden locations or metal-enclosed areas. 

Carbon-fibre and metal composite materials may result in significant multi-path due to signal reflections~\cite{Panitz10,WAIC_spc}. Hence, each wireless link might have high bit error rates due to multi-path signals in metal-enclosed areas. The metal composite materials are extremely hard to propagate radio waves. Furthermore, the RF signals are substantially attenuated through absorption and shadowing effect due to some interior structures, passengers, baggage, and cargo. Although the metal-enclosed attenuation seriously increases the path loss, it gives potentials to efficiently share and reuse the communication resources among WAIC transceivers inside the airframe. If the RF isolation between different compartments is sufficient, multiple networks could operate at the same radio frequency. 

Outside WAIC transceivers usually experience lower path loss compared to applications inside the airframe. However, the safety-critical sensors and actuators installed outside the airframe are vulnerable to external interference from other aircraft or jamming attack. 


\subsection{Operation-Dependent Heterogeneous Applications}  
WAIC systems are complex and support heterogeneous classes of functionalities, as we have discussed in Section~\ref{sec:waic}. While some functions of aircraft are very safety-critical, most of them are less critical based on Tables~\ref{tab:DO178} and~\ref{tab:class}. WAIC must support both safety-critical and non safety-critical functionalities since non safety-critical ones can result in significant performance losses or damages to the aircraft  in the long-term. 


Most of the WAIC applications continuously operate during all flight phases, while some applications are only active for limited periods of time depending on the aircraft operations. For instance, slats and flaps of Fig.~\ref{fig:RefAir} are mainly used for low speed control during takeoffs and landings~\cite{Schmitt98}, while cameras mounted outside the airframe require relatively low data rate during aircraft parking. The WAIC systems must adapt its resource allocations based on these operational-dependent demands. 

The aircraft operations could affect the radio propagation property since some parts of aircraft change position. During landing and takeoff, it is possible to provide the reliable communication link between expanded landing gear sensors and gateway installed outside the airframe. However, the path loss significantly increases when the gears are closed during the in-flight mode.


\subsection{\Blue{Natural Disturbing Factors}}
\Blue{WAIC system must support the time-critical operation, even in the harsh environment due to various pressure, temperature, and humidity and unexpected electrometeors and solar activities. While the wire shielding protects wired data transmission against outer distortions, wireless communication systems are inherently vulnerable to natural disturbances. Several natural disturbances possibly affect the signal propagation characteristics of the WAIC network~\cite{alt_1,alt_2,rain_1}. The critical natural disturbances on the signal attenuation are classified into four categories, namely, atmospheric gases, hydrometeors, electrometeors, and solar activities~\cite{Aglargoz14}. Several gases in the atmosphere cause signal degradation on a transmitting wave~\cite{alt_1, alt_2}. Hydrometeors include all atmospheric water-formed particles, such as rain and cloud, that affect the wireless link as well~\cite{rain_1,rain_2}. Furthermore, the electrometeors such as lightning and electrical field and solar activities also severely affect the radio propagation characteristic~\cite{Aglargoz14,Inc18}.}



\subsection{Coexistence and Interference Issue}  
Since the frequency band between $4.2-4.4$~GHz does not overlap with the ISM band, there is no serious concern with typical interference issues from existing wireless communication technologies. However, each aircraft still has to share the spectrum resources of $4.2-4.4$~GHz with other aircrafts. Since aircrafts are widely spaced apart during in-flight mode to avoid the mid-air collision, the interference factor becomes negligible. Indeed, two aircraft are vertically separated by at least 300~m during the in-flight mode~\cite{Park14,Park14_TIT}. 

However, the interference problem becomes severe when many aircraft are very closely located at the airport during parking or taxiing. On the ground, outside WAIC transceivers may cause strong interference to other aircraft in contrast to inside ones. Many critical sensors and actuators outside the airframe are essentially vulnerable to co-channel interference and attacks since the airframe does not provide any protection. Hence, the interference between aircraft must be controlled to allow coexistence. Different diversity domains such as time, frequency, and space are some of the obvious countermeasures to handle interference~\cite{Yang10}. However, some advanced techniques such as interference cancellation and software defined radios can be applied as well. 


\begin{table*}[t]
\centering{
\begin{tabular}{ |c|c|c|c|c| } 
 \hline
{}  & {} & \multicolumn{2}{c|}{Low Data Rate}  & High Data Rate \\  
\cline{3-5}
{}  & {} & 802.15.4-based Solutions & Bluetooth & 802.11-based Solutions\\ 
\hline
\multirow{2}{*}{Operational Requirements} & Reliability and Latency Requirements & $\odot$ & $-$ & $-$\\ 
\cline{2-5}
{}  & Security & $\odot$ & $\odot$ & $\odot$ \\
 \hline
\multirow{4}{*}{Design Constraints} & Structural Environment & $-$ & $-$ & $-$ \\
\cline{2-5}
{}  & Operation-Dependent Heterogeneous Applications & $\odot$ & $-$ & $\odot$ \\ 
\cline{2-5}
{} & Natural Disturbing Factors  & $-$ & $-$ & $-$ \\
\cline{2-5}
{}  & Coexistence and Interference Issue  & $\odot$ & $\odot$ & $\odot$ \\
 \hline
\multirow{2}{*}{Monetary Cost} & Energy Efficiency & $\oplus$ & $\oplus$ & $-$ \\
\cline{2-5}
{} & System Integration & $\oplus$ & $\oplus$ & $\oplus$ \\
 \hline
\end{tabular}
\caption{\Blue{Research maturity levels of candidate wireless standards, namely, 802.15.4-based solutions and Bluetooth for low data rate and 802.11-based solutions for high data rate. By considering the specific challenge, each research field of various standards is marked as $(\oplus, \odot, -)$ from most to least mature one.}}\label{tab:lev_std}}
\end{table*}

\subsection{Energy Efficiency}
Low data rate applications may use battery operated nodes to further reduce the deployment and maintenance cost of WAICs. Note that all energy-constrained WAIC nodes must operate flawlessly during a long flight (below 20 hours). Even though the true wireless device, operated on battery power, gives the great benefit for the WAIC operation, the energy resource must be carefully chosen~\cite{Korber07}. For instance, some devices do not have any line power supply nearby such as pneumatic actuators, whereas other nodes have access to it~\cite{Gud14_struct}. Furthermore, the aircraft operators may need to maintain various batteries of WAIC nodes to replace consumed batteries. Hence, the life-cycle cost of battery operated wireless nodes can be more expensive compared to nodes that only communicate via the wireless medium.

Energy harvesting techniques harmoniously compensate the limitation of batteries since aircraft generates consistent vibration and heat when they are operating~\cite{Xu16}. For instance, the vibration of the aircraft environment is a good energy source that can be effectively harvested either by electromagnetic induction mechanism or piezoelectric energy conversion mechanism~\cite{Dilhac14,Shin14_1}. By using moving magnets or coils, vibrational magnetic generators can produce energy in the range of microwatts to milliwatt depending on the microelectromechanical system. Piezoelectric energy harvesting devices also generate the output power density of 100 and 330~$\mu$W/cm$^3$~\cite{Anton07, Shin14_2}. In space shuttles, energy harvesting technology has been successfully applied to reduce the maintenance costs of wireless sensors~\cite{invocon2}. 



\subsection{System Integration}
WAIC systems must not only meet the specific requirements of certification, but also provide interoperability with existing avionics infrastructures. In Fig.~\ref{fig:waic}, the gateway is one of most critical points to efficiently integrate WAIC systems. However, since a small number of gateway vendors only propose proprietary solutions,  the efficient integration to existing infrastructure is complicated. It is essential to standardize the WAIC integration into various avionics fieldbuses to achieve simple deployment, low installation cost, and maintenance.



\section{Candidate Wireless Solutions for WAIC} \label{sec:existing}   
Developing a WAIC system from scratch is inefficient since a large number of wireless standards already successfully used in various commercial applications. Hence, we investigate the applicability of existing standards to support the heterogeneous services of both low data rate and high data rate applications. For the low data rate applications, IEEE 802.15.4~\cite{154} is one of the strong candidates because it provides sufficient range of communication, energy efficiency, scalability, and flexibility. Most industrial low rate wireless standards focus on the real-time communication based on the 802.15.4 physical (PHY) layer for a large-scale open-loop sensing and monitoring of the non-critical process automation~\cite{Kumar14,Palattella13}. Some experimental results~\cite{Kadri12, Delsing10, Bertocco08} show the applicability of the IEEE 802.15.4 PHY layer for harsh industrial environments. Although it is not originally designed for mission-critical applications, Bluetooth is also considered and reviewed as a candidate technology for WAIC systems. The main reasons are the huge ecosystem of existing devices and the interoperability guaranteed by a full protocol stack certification~\cite{Bluetooth}. 

Regarding high data rate applications, the IEEE 802.11 standard is a good basic technology since it offers the high throughput with a number of possible modulation and coding options~\cite{11b, 11n}. WLAN can be used to deliver non-traditional variables such as voice and video, and to efficiently replace the existing high data rate fieldbus such as AFDX~\cite{A664} and AS6802~\cite{6802}. While enabling high data rate wireless communications was a significant achievement in WLAN network, they do not guarantee any deterministic network performance in terms of availability, reliability, and latency. Hence, the traditional solutions are not appropriate for the real-time control systems~\cite{Pellegrini06}. 

\begin{figure*}[t]
  \centering
  \includegraphics[width = 1.9 \columnwidth]{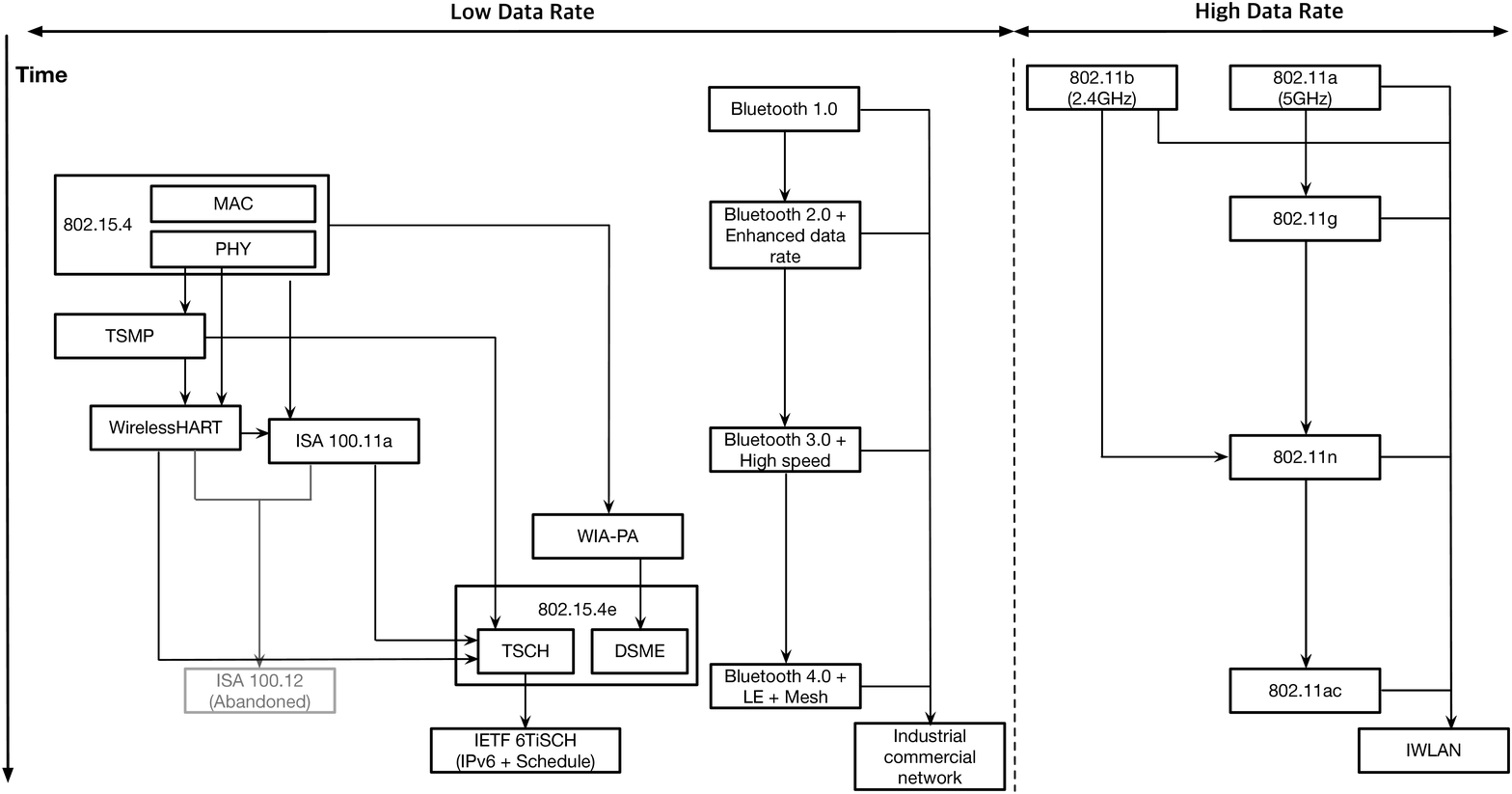}  %
   \caption{Historical evolution of 802.15.4-based, Bluetooth-based, 802.11-based industrial wireless networks for low and high data rate applications.}
   \label{fig:evo}
\end{figure*}

This section presents the brief overview of most related 802.15.4-based industrial standards, the Bluetooth standard, and 802.11 standards. \Blue{Table~\ref{tab:lev_std} presents the research maturity levels of candidate wireless standards along the technical challenges. By considering the specific challenge, each research field of various standards is marked as $(\oplus, \odot, -)$ from most to least mature one.} Table~\ref{tab:existing} summarizes the main features of the PHY and MAC layers of IEEE 802.15.4, Bluetooth, and IEEE 802.11. Fig.~\ref{fig:evo} describes the historical evolution of 802.15.4-based, Bluetooth-based, and 802.11-based standards for industrial contexts.


\begin{table*}[t]
\centering{
\begin{tabular}{ |c|c|c|c|c|c| } 
 \hline
 PHY/MAC  & \multirow{2}{*}{IEEE 802.15.4} & \multirow{2}{*}{Bluetooth} & \multirow{2}{*}{IEEE 802.11ac} & \multirow{2}{*}{IEEE 802.11ax}  & \multirow{2}{*}{IEEE 802.11ay}\\  
 Technology & & & & & \\
 \hline
Range & $<10$m & $<200$m & $<200$m & $<200$m & $<60$m, Extension $300-500$m\\ 
 \hline
\# of antennas  & 1 & 1 & 4 & 8 & 4\\
 \hline
Packet error rate & 0.01 & 0.1 & 0.1 & 0.1 & 0.1\\ 
 \hline
\multirow{2}{*}{FEC}  &  Convolutional code & \multirow{2}{*}{24-bit CRC} & Convolutional code & Convolutional code  & \multirow{2}{*}{LDPC, STBC} \\ & No FEC, RSC &  & LDPC, STBC & LDPC, STBC & \\
 \hline
\multirow{2}{*}{Frequency}  & 780/868/915/950~MHz,  & \multirow{2}{*}{2.4~GHz} & \multirow{2}{*}{2.4~GHz, 5~GHz} & \multirow{2}{*}{5~GHz} & \multirow{2}{*}{60~GHz} \\ &  2.4~GHz &&&&\\
 \hline
Bandwidth & 200~kHz-5~MHz & 2~MHz & 20~MHz-40~MHz & 20~MHz-160~MHz & 2.16~GHz-8.64~GHz \\ 
 \hline
Peak data rate & $250$~kb/s & $2$~Mbit/s & $< 6.9$~Gbit/s & $<9.6$~Gbit/s & $20-40$~Gbit/s\\ 
 \hline
Signaling & DSSS & FHSS & OFDM & OFDM & OFDM\\ 
 \hline
Modulation & Chirp­SK/FSK/PSK/ASK & GFSK & Up to 256­-QAM & Up to 1024­-QAM & Up to 256-QAM \\ 
 \hline
Medium access & TDMA and CSMA & Aloha and FDMA & TDMA and CSMA & CSMA and OFDMA & TDMA and CSMA \\ 
 \hline
\end{tabular}
\caption{PHY and MAC layer features of IEEE 802.15.4~\cite{154}, Bluetooth~\cite{Bluetooth}, IEEE 802.11ac~\cite{11ac}, IEEE 802.11ax~\cite{11ax}, and IEEE 802.11ay~\cite{11ay}.}\label{tab:existing}}
\end{table*}

\subsection{Low Date Rate}
Since the baseline IEEE 802.15.4~\cite{154} has some deficiencies to support the reliable real-time communications~\cite{Park18, Wang16}, various industrial alliances including the HART communication foundation~\cite{whart}, the International Society of Automation~\cite{isa100}, and the Chinese Industrial Wireless Alliance~\cite{wia} have established WirelessHART~\cite{whart}, ISA100.11a~\cite{isa100}, and Wireless Networks for Industrial Automation-Process Automation (WIA-PA)~\cite{iec_pa} standards, respectively. All these standards adopt the complete IEEE 802.15.4 PHY layer. WirelessHART and ISA 100.11a propose a new Medium Access Control (MAC) layer combining Time Division Multiple Access (TDMA) and channel hopping techniques, whereas the WIA-PA standard renovates the existing IEEE 802.15.4 MAC layer. Main targeting applications of these standards are monitoring and control of process automation and equipment which has relaxed requirements with respect to the ones of critical WAIC applications~\cite{Park17}. 

This section briefly discusses WirelessHART, ISA100.11a, WIA-PA. We refer the papers~\cite{Kumar14, Wang16, Liang11} for extensive analysis and summary of industrial wireless standards for low data rate applications. Table~\ref{tab:wsn} summarizes main technical features of WirelessHART, ISA 100.11a, and WIA-PA. As an alternative candidate for low data rate WAIC applications outside the 802.15.4-based solutions, the Bluetooth technology is also discussed in this section.

\subsubsection{WirelessHART}
WirelessHART is a wireless communication standard designed for industrial process automation~\cite{whart}. This standard operates at $2.4$~GHz frequency and uses 15 different channels on top of the IEEE 802.11 PHY layer~\cite{Chen10}. WirelessHART combines the advantages of both TDMA and channel hopping in order to efficiently handle the fading and noise effect. The main strategies of the MAC and network layers are based on Time Synchronized Mesh Protocol (TSMP) designed by Dust Networks~\cite{TSMP_dust}. To migrate the multi-path fading, channel hopping allows data transmission at different frequencies even for the periodically allocated time slots. Another feature of WirelessHART to increase reliability is graph routing~\cite{Lu16}. A graph is a set of paths connecting network nodes~\cite{Song08}. In contrast to typical routing techniques, the graph routing uses a specific graph ID instead of node ID. Because of its knowledge of the entire set of route information, the network manager builds the paths and provides the graph information to each node. 

The main elements of a WirelessHART network are field devices, a gateway, Access Points (AP), and a network manager~\cite{Chen10}. The field devices include sensors attached to the industrial plants, in order to provide data or utilize data from other devices. Field devices have full routing capability, whereas AP only connects field devices to the gateway. The centralized network manager configures the network and manages the resource allocation between field devices.

\begin{table*}[t]
\centering
\begin{tabular}{ |c|c|c|c| } 
 \hline
 Technology & WirelessHART & ISA100.11a & WIA-PA \\ 
 \hline
Number of channels & 15 (2.4~GHz Band) & 16 (2.4~GHz Band) &  16 (2.4~GHz Band)\\ 
 \hline
Superframe & Collection of timeslots & Collection of timeslots & 802.15.4 Superframe\\ 
 \hline
Multiple access scheme & TDMA and CSMA & TDMA and CSMA & TDMA and CSMA\\ 
 \hline
Timeslot duration & 10ms & Flexible and configurable  & Configurable \\ 
 \hline
Routing & Source/Graph & Source/Graph & Static routing\\ 
 \hline
Network topology & Star, Mesh & Star, Mesh & Mesh + Star\\ 
 \hline
Clock tolerance & 10ppm & 10ppm or Relaxed requirement & Relaxed requirement\\ 
 \hline
Resource allocation & Centralized & Centralized & Centralized + Distributed\\ 
 \hline
\end{tabular}
\caption{Comparison of technical features of WirelessHART~\cite{whart}, ISA 100.11a~\cite{isa100}, and WIA-PA~\cite{wia} for low data rate industrial applications.}\label{tab:wsn}
\end{table*}

\subsubsection{ISA100.11a}
Similar to WirelessHART, ISA100.11a mainly relies on TDMA with channel hopping techniques at the data link layer~\cite{isa100}. To decrease interference effects, ISA100.11a provides more options of channel hopping, including slow hopping, fast hopping, and mixed hopping. While WirelessHART mainly supports integration with existing wired HART, ISA100.11a provides IPv6 compatibility to connect the Internet as well as the inter-operability with legacy protocols such as Modbus, Profibus, and wired HART~\cite{Dujovne14}.

ISA100.11a defines two main classes of devices, namely, field devices and infrastructure devices~\cite{isa100}. Compared to WirelessHART, ISA categorizes the field devices into three classes: routing, I/O, and handheld devices. I/O devices are reduced function devices to provide data or utilize data without supporting the routing capability. Infrastructure devices include backbone routers, gateways, and system managers. Backbone routers act as a proxy since they route the data packets to their destinations through the backbone network.

\subsubsection{WIA-PIA}
In contrast to WirelessHART and ISA100.11a, WIA-PA is a fully compatible standard with the existing IEEE 802.15.4 standard, for industrial process automation~\cite{wia, iec_pa}. The MAC layer is enhanced by the combination of Carrier Sense Multiple Access (CSMA), TDMA, and FDMA schemes built on the IEEE 802.15.4 standard. It supports various legacy protocols such as wired HART, Profibus, Modbus, and WirelessHART.

By considering the basic IEEE 802.15.4 topology, WIA-PA combines hierarchical clustering with a mesh topology between cluster heads. The main elements of a WIA-PA network are routing device, gateway device, and field device. As cluster heads, routing devices construct and monitor their clusters and route the collected data packets to the backbone network. In each cluster, field devices called cluster members, measure plant states and directly transmit it to their cluster heads without having the routing capability. To avoid a single point failure, the standard recommends a redundant cluster head as a backup~\cite{Liang11}.

\subsubsection{Bluetooth Low Energy}
\Blue{The major usage of Bluetooth is the wireless connection between a mobile phone and a headset. Bluetooth Low Energy (BLE)~\cite{Bluetooth}, released in 2010 as part of Bluetooth 4 specification, has been standardized to expand the ecosystem of Bluetooth to the Internet of Things. The focus is primarily in the smart home sector, but it could possibly impact the industrial applications because of the huge ecosystem of compatible devices (e.g., smartphones)~\cite{BLE_ind}. BLE provides two data transfer modes, namely, connection-oriented and connectionless modes. In the connection-oriented mode, devices negotiate dedicated resources for data communications. On the other hand, random access channels are used to directly exchange data in the connectionless mode. A modified connectionless approach has been introduced in Bluetooth~5 to enhance advertising capabilities. A tradeoff between reliability, delay, and sensor lifetime is analyzed depending on the data transmission mode~\cite{PG17}. Furthermore, Bluetooth has recently introduced mesh networking capabilities to handle networks of hundreds of devices with low latency and robustness~\cite{PGWP}.}


\subsubsection{Discussion}
The low-cost and low-power of these wireless standards make it appealing for low data rate WAIC applications~\cite{Wang16, Liang11}. Some simulation studies are carried out to evaluate the performance of these standards: WirelessHART~\cite{Willig05, Tezcan07, Zhang07, Hull04, Mahmood15}, ISA100.11a~\cite{Stann03, Wan04, Akerberg11, Dominicis09}, WIA-PA~\cite{Pietro07, Melodia07}, BLE~\cite{Rondon17}, but they have different parameters or tools over various environments. Hence, it is hard to derive common conclusions due to the lack of quantitative performance comparison in a uniform simulation environment.

WirelessHART~\cite{whart} and ISA100.11a~\cite{isa100} are two dominating standards in a real industrial market. Both standards are very similar, but ISA100.11a provides more flexible techniques such as flexible timeslot length, various hopping patterns, and distributed resource allocation. However, these appealing options lead to the practical challenges such as protocol complexity and interoperability. Hence, developing ISA100.11a-based system is more challenging than the one using WirelessHART. A subcommittee named ISA100.12 investigates the convergence between these standards~\cite{Hayashi09,Petersen11}. However, the technical convergence has not been well agreed and the subcommittee has been disbanded.

The IEEE 802.15.4e standard has also enhanced the IEEE 802.15.4 MAC layer to support the industrial process applications. Two major modes of the standard are Time Slotted Channel Hopping (TSCH) based on WirelessHART and ISA100.11a, and deterministic and synchronous multi-channel extension based on WIA-PA. Since these modes do not interoperate, the operator needs to select one mode in a network. 

In~\cite{Park17}, we evaluate the feasibility of different industrial wireless networks such as WirelessHART~\cite{whart}, IEEE 802.15.4e~\cite{154e}, and wireless interface for sensors and actuator~\cite{wisa} with respect to the flight safety requirements of Table~\ref{tab:DO178}. While enabling real-time wireless communications is a significant achievement in these industrial communications, availability and reliability requirements of the flight certification exceed what current networks can offer~\cite{Park17}. Hence, some improvements are still recommended for safety-critical WAIC applications. In~\cite{Rondon17}, an analysis of the BLE performance for time-critical industrial applications is presented, relying on various retransmission schemes. Although high reliability and low latency constraints can be achieved separately, the simultaneous fulfilment of all requirements for flight safety is not achieved.


\subsection{High Data Rate}
Recently, WLAN-based industrial wireless networks become very popular due to its technical maturity for the high data rate applications~\cite{iwlan1,Cena10,Tramarin15}. IEEE 802.11 is a set of PHY and MAC specifications to implement WLAN communication. The PHY layer defines a series of transmission schemes including modulation and coding techniques for wireless communications. In the MAC layer, two basic access schemes are the contention-based Distributed Coordination Function (DCF) and the optional contention free Point Coordination Function (PCF)~\cite{Hiertz10}. The DCF scheme is the default access scheme of IEEE 802.11 systems based on Carrier Sense Multiple Access with Collision Avoidance (CSMA/CA). When centralized PCF is activated, the wireless communication is organized with temporal windows called superframes. Each superframe period is further divided into a Contention Period (CP) using DCF and a Contention-Free Period (CFP) using PCF. Then, an AP periodically polls high priority nodes and allocates the transmitting slots to them for time-critical applications. It only operates under the centralized infrastructure mode. However, PCF is rarely supported by manufacturers in practice.

This section briefly introduces the essential technical features of main WLAN standards such as traditional 802.11 a/b/e/g, relatively new 802.11n/ac, and upcoming 802.11ax and 802.11ay for the high data rate applications.

\subsubsection{802.11a/b/e/g}
As the first widely accepted standard, IEEE 802.11b achieves the $11$~Mbit/s data rate at $2$~GHz using Direct Sequence Spread Spectrum (DSSS) modulation scheme~\cite{11b}. In the $5$~GHz band, IEEE 802.11a provides the $54$~Mbit/s data rate based on the Orthogonal Frequency-Division Multiplexing (OFDM) technique~\cite{11a}. In 2003, IEEE 802.11g integrated OFDM transmission scheme with IEEE 802.11b PHY layer to improve the data rate to $54$~Mbit/s in the $2.4$~GHz band. 

The general WLAN standards such as 802.11a/b/g are widely used for various applications, but they do not assure any deterministic reliability, latency, and throughput performance for time-critical applications~\cite{Zhu04,Hiertz10}. To serve different Quality of Service (QoS) requirements of various applications, the IEEE 802.11e standard introduces a new adaptive framework called Hybrid Coordination Function (HCF)~\cite{11e}. HCF modifies the original DCF and PCF to support both new QoS features of time-critical applications and non-QoS features of traditional applications within a network. For instance, IEEE 802.11e is used to provide the low latency communication for the factory automation system~\cite{Cena10}.

\subsubsection{802.11n/ac}
In 2009, the IEEE 802.11n standard considerably enhanced the data rate up to $600$~Mbit/s over $54$~Mbit/s of the earlier 802.11a/g~\cite{11n}. Three core techniques of the PHY and MAC layers are Multiple Input Multiple Output (MIMO) antennas, channel bonding, and frame aggregation. Most 802.11n improvements rely on the MIMO technique since it allows to receive or transmit simultaneously using multiple antennas~\cite{Zheng15}. The spatial division multiplexing of MIMO spatially multiplexes different data streams to increase data throughput. For each spatial stream, both the transmitter and the receiver require a discrete antenna. On the other hand, multiple antennas could send redundant information to improve the reliability. A useful technique for this purpose is Space Time Block Coding (STBC), whereby multiple copies of data are sent over different antennas to enhance the reliability at the expense of the bit rate.

The channel bonding of the PHY layer basically combines adjacent channels to increase the throughput. Two adjacent $20$~MHz channels are combined to form a single $40$~MHz channel to double the throughput. Furthermore, IEEE 802.11n aggregates multiple frames into a single data unit to reduce the interframe spaces and headers. By considering major enhancements, recent industrial APs use the IEEE 802.11n standard to support the high reliable and high data rate demands~\cite{moxa,eki6311}.

As the most recent development regarding the very high throughput amendment, IEEE 802.11ac offers a theoretical maximum rate of 6.93~Gbit/s~\cite{11ac}. Compared to 802.11n, it supports higher-order modulation, wider channels, more MIMO spatial streams, and downlink Multi-user MIMO (MU-MIMO) in the $5$~GHz band. In contrast to the single user-MIMO of 802.11n, the downlink MU-MIMO supports four simultaneous downlinks connected to the same AP. 

\subsubsection{802.11ax}
IEEE 802.11ax~\cite{11ax}, also marketed as 802.11 High Efficiency Wireless or Wi-Fi 6 by the Wi-Fi Alliance, is expected to be standardized in 2020. 802.11ax is designed to operate in all band spectrums between $1$ and $7$~GHz in addition to the already existing $2.4$ and $5$~GHz bands. The amendment is designed to improve performance of WLAN deployments in dense scenarios, characterized by large number of APs and associated nodes deployed in geographical limited region, e.g. a stadium or an airport. The amendment targets four times throughput improvement with respect to 802.11n and 802.11ac. However, the practical throughput is still lower than 10~Gbit/s~\cite{11ax}. 


\subsubsection{802.11ay}
\Blue{IEEE 802.11ay~\cite{11ay} enhances 802.11ad~\cite{11ad_mm} in the 60 GHz millimeter waveband.  The main goals of this standard are to achieve a throughput of at least 20~Gbit/s and to increase the transmission range and reliability. The current draft offers a $20-40$~Gbit/s transmission rate and an extended $300-500$~m transmission range. 802.11ay uses a maximum bandwidth of 8.64~GHz by bonding four of 2.16~GHz bandwidth of 802.11ad. Similar to previous WLAN standards~\cite{11ax,11n}, 802.11ay includes channel bonding, channel aggregation, and MU-MIMO technologies. The target applications are to replace indoor Ethernet and other cables and to provide outdoor backhaul connectivity.}






\begin{table*}[t]
\fontsize{7.8pt}{10pt} \selectfont
\centering
{
\begin{tabular}{ |c|c|c|c|c|c|c|c|c| } 
 \hline
{}  & \multicolumn{2}{c|}{Operational Requirements} & \multicolumn{4}{c|}{Design Constraints} & \multicolumn{2}{c|}{Monetary cost} \\  
\cline{2-9}
{}  & Reliability and   & Security  & Structural &  Operation-   & Natural  & Coexistence and  & Energy & System \\ 
{}  & Latency  & {}  & Environment & Dependent  &  Disturbing & Interference Issue  & Efficiency & Integration  \\ 
{}  & Requirements & {}  & {} & Heterogeneous  & Factors & {} & {} & {}  \\ 
{}  & {} & {}  & {} & Applications  & {} & {} & {} & {}  \\ 
 \hline
Wireless Channel & {$-$} & {} & {$\odot$} & {$-$} & {$\odot$} & {} & {} & {}\\ 
\hline
Spectrum Division & {} & {} & {$\oplus$} & {} & {} & {$\oplus$} & {} & {$\oplus$}\\ 
 \hline
Network Architecture & {$\odot$} & {} & {$\odot$} & {$\odot$} & {} & {} & {} & {}\\ 
 \hline
Link Adaptive Scheme & {$\odot$} & {} & {} & {} & {} & {} & {} & {}\\ 
 \hline
Hybrid Multiple Access & {$\odot$} & {} & {} & {} & {} & {} & {$\oplus$} & {}\\ 
 \hline
Resource Allocation & {$\odot$} & {} & {$\odot$} & {$\odot$} & {} & {} & {$\oplus$} & {}\\ 
 \hline
Time Standard and & \multirow{2}{*}{$\oplus$} & {} & {} & {} & {} & {} & \multirow{2}{*}{$\oplus$} & {}\\ 
Synchronization & {} & {} & {} & {} & {} & {} & {} & {}\\ 
 \hline
Multi-hop Routing & {$\odot$} & {} & {$\odot$} & {} & {} & {} & {} & {}\\ 
 \hline
Machine Learning-based  & \multirow{2}{*}{$-$} & {} & \multirow{2}{*}{$-$} & \multirow{2}{*}{$-$} & {} & {} & \multirow{2}{*}{$-$} & {}\\ 
Parameter Management & {} & {} & {} & {} & {} & {} & {} & {}\\ 
 \hline
Effective Security & {} & {$\odot$} & {} & {} & {} & {} & {} & {}\\ 
 \hline
Joint Design of Control  & \multirow{2}{*}{$-$} & {} & \multirow{2}{*}{$-$} & \multirow{2}{*}{$-$} & {} & {} & {} & {}\\ 
and Communication & {} & {} & {} & {} & {} & {} & {} & {}\\ 
 \hline
Fault-Detection and & \multirow{2}{*}{$\odot$} & \multirow{2}{*}{$-$} & {} & {} & \multirow{2}{*}{$-$} & {} & {} & {}\\ 
Fault-Tolerant Design & {} & {} & {} & {} & {} & {} & {} & {}\\ 
 \hline
Prediction and   & \multirow{3}{*}{$-$} & {} & {} & {} & \multirow{3}{*}{$-$} & {} & {} & {}\\ 
Adaptation for & {} & {} & {} & {} & {} & {} & {} & {}\\ 
Natural Disturbances  & {} & {} & {} & {} & {} & {} & {} & {}\\ 
 \hline
Effective Directional & \multirow{2}{*}{$\odot$} & {} & \multirow{2}{*}{$\odot$} & {} & {} & {} & {} & {}\\ 
Antenna and MIMO& {} & {} & {} & {} & {} & {} & {} & {}\\  
 \hline
Effective Integration & {} & {} & {} & {} & {} & {} & {} & {$\odot$}\\ 
 \hline
Power Delivery and & {} & {} & \multirow{3}{*}{$\odot$} & {} & {} & {} & \multirow{3}{*}{$\odot$} & \multirow{3}{*}{$\odot$}\\ 
Data Transmission & {} & {} & {} & {} & {} & {} & {} & {}\\ 
Through Metals & {} & {} & {} & {} & {} & {} & {} & {}\\ 
 \hline
\end{tabular}
\caption{\Blue{Research maturity levels of different technical schemes along various challenges. By considering the specific challenge, each research field of the technical scheme is marked as $(\oplus, \odot, -)$ from most to least mature one. Note that the blank denotes the irrelevance or weak contribution of the technique corresponding to the challenge.}}\label{tab:lev_cha}}
\end{table*}

\subsubsection{Discussion}
WLAN is an excellent network infrastructure for high data rate applications with readily available off-the-shelf products. However, the IEEE 802.11 technology is not yet capable of the real-time transmission and high data rate demands of WAIC applications due to some fundamental limitations on throughput, reliability, and latency~\cite{Zhu04,iwlan1}. The main reason is that the default contention-based DCF scheme does not inherently provide any deterministic performance. 

PCF is the preferable solution since the network resources can be deterministically assigned to ensure the transmission of data within a specific period. However, PCF does not satisfy the strict real-time requirements of critical industrial applications since the transmission periods are in the range of several hundred milliseconds~\cite{Santandrea06}. To deal with this problem, The Industrial Wireless LAN (IWLAN) technology extends the IEEE 802.11 standard using industrial PCF protocol to meet the real-time and redundancy requirements~\cite{iwlan1, iwlan2}. In this protocol, a master simply uses a polling scheme to manage the sequence of transmissions. However, the throughput of IWLAN is low as the protocol assigns each time slot to all associated nodes.

Since MIMO is one of the most successful techniques of 802.11n/ac/ax/ay, WAIC systems can adapt MIMO to enhance both the throughput and the reliability of the network. In particular, MU-MIMO could lead a significant change of traditional WLAN networks since multiple nodes are allowed to simultaneously use the same channel. Moreover, recent 802.11ax supports both downlink and uplink MU-MIMO compared to the downlink one of 802.11ac~\cite{11ax}. Uplink MU-MIMO technique is particularly useful for WAICs since it allows to transmit sensing values from multiple nodes to the single AP at the same channel. 




\section{Fundamental Techniques}\label{sec:tech}  
\Blue{Despite of the great advancement of current industrial wireless networks, no existing industrial standard can be used exclusively to meet the safety requirements of WAICs,  as illustrated in Table~\ref{tab:lev_std}. This section discusses most promising techniques based on existing industrial wireless networks of Section~\ref{sec:existing} to handle the complex demands of Section~\ref{sec:waic} and the major technical challenges of Section~\ref{sec:cha}. In particular, this section can help the readers to select some interesting unsolved issues by considering recent development of joint design of control and communication~\cite{Park18}, machine learning~\cite{Goodfellow16}, and edge computing~\cite{Chen19}. Table~\ref{tab:lev_cha} summarizes the research maturity levels of different technical schemes along various challenges. By considering the specific challenge, each research field of the technical scheme is marked as $(\oplus, \odot, -)$ from most to least mature one. Note that the blank denotes the irrelevance or weak contribution of the technical scheme corresponding to the challenge.}



\subsection{Wireless Channel}

The complex aircraft structure consisting with metallic ceilings, metallic joists, as well as moving parts affects the scattering and reflection characteristics of the radio channel~\cite{WAIC_obj}. The PHY and MAC layers of WAIC system need to adapt their parameters dependent on the large-scale and small-scale fading effect of wireless channels. 

The ITU-R report~\cite{WAIC_spc} provides the signal propagation model of a typical wide-body aircraft (an aircraft with two aisles) based on experimental measurements. By combining analytical and empirical schemes, a log-normal shadowing path loss model was used to model the wireless channel in various positions of the aircraft. Various sets of propagation measurements are used to develop the log-distance path model to capture the large-scale fading. Table~\ref{tab:group} presents the set of test locations and path loss exponents~\cite{WAIC_spc}. These realistic link models can be used for simulation-based studies. The range of the path loss exponent is between 1.51 and 3.46. For group E, the calculated path loss exponent is less than the one of free space, since the cabin acts as a resonant cavity. When test pairs are separated by deck floor or fuselage, the path loss exponents are between 2.12 and 2.49, due to a NLOS path. For group D, the cabin windows provide some near-LOS component. The largest path loss exponent 3.46 is obtained when both transmitter and receiver are separated by cabin monuments such as lavatories, galleys, etc., for group B.

\begin{table*}[t]
\centering
\begin{tabular}{ |c|c|c|c| } 
 \hline
Group & Group name & Path loss exponent $(n)$ & Description\\ 
 \hline
A & Intra-Cabin \& Intra-Flight Deck & $2$ & Both TX and RX are in the same cabin area\\ 
 \hline
B & Inter-Cabin & $3.46$ & TX and RX are separated by cabin monuments \\ 
 \hline
\multirow{2}{*}{C} & Inter-Cabin-to-Lower Lobe \& & \multirow{2}{*}{$2.49$} & \multirow{2}{*}{TX and RX are separated by the main deck floor}\\ 
{} & Inter-Cabin-to-Flight Deck & {} & {} \\ 
 \hline
D & Inter-Cabin-to-Exterior & $2.12$ & TX and RX are separated by the fuselage \\ 
 \hline
\multirow{2}{*}{E} & Inter-Cabin-to-Landing Gear \&  & \multirow{2}{*}{$1.51$} & \multirow{2}{*}{TX and RX are separated by the airframe} \\ 
{} & Inter-Lower-Lobe to Exterior & {} & {} \\ 
\hline
F & Inter-Exterior & $2.31$ & Both TX and RX are exterior of the aircraft fuselage \\ 
 \hline
\end{tabular}
\caption{Wireless channel model parameters of log-distance path loss model for each group of test pairs between transmitters and receivers~\cite{WAIC_spc}.} \label{tab:group} 
\end{table*}

\begin{figure}[t]
  \centering
  \psfrag{x}[][]{\footnotesize{Distance (m)}}
  \psfrag{y}[][]{\footnotesize{Path loss (dB)}}  
  \includegraphics[width = 0.9 \columnwidth]{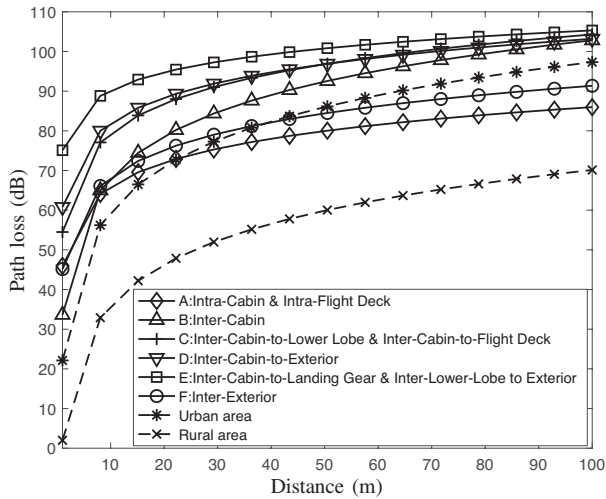}  %
   \caption{Path loss as a function of different distances between transmitter and receiver with all groups (A-F) of Table.~\ref{tab:group} and urban and rural areas of general cellular communications~\cite{3GPP}.}
   \label{fig:loss}
\end{figure}

Fig.~\ref{fig:loss} shows the path loss of all groups (A-F) of Table~\ref{tab:group} and general cases of urban and rural ares of cellular communications~\cite{3GPP} as a function of various distances between the transmitter and the receiver at $4.2$~GHz. Note that we include the path loss of the general cellular communication as the reference to compare. We clearly observe the significant path loss of various test pairs with respect to the one of the rural area. In fact, the group E gives the largest path loss since the transmitter and the receiver are isolated by the fuselage. To support the reliable communication, a large fading margin must be chosen to overcome a deep fade due to possibly moving carts and passengers even in the cabin. Spread spectrum modulation can be also used to improve the availability against deep fading. 

The amount of interference between compartments mainly depends on the structure and material of bulkheads. Based on the channel measurements, it is possible to estimate the interference between compartments~\cite{WAIC_spc}. Then, we may reuse the channels whilst meeting the minimum signal-to-interference ratio for non-interfering regions. Since outside WAIC environment generally has good channel propagation property, the frequency reuse outside the airframe is not recommended on a single aircraft.

In contrast to the extensive channel measurements of the large-scale fading~\cite{WAIC_spc}, the small-scale fading effect is not well established for aircraft. To characterize the small-scale fading effects, field tests have been conducted to measure the received power distribution in the light aircraft~\cite{Park19}. We used a software-defined radio platform and designed monopole antennas in order to make them compliant with the $4.2-4.4$~GHz frequency band of WAIC. The communication link was observed between several points such as cockpit, engine room, rear room, and wing spots. Note that different compartment environments are filled with electronic devices and mechanic components in practice. The root mean square delay between measured points for NLOS transmission is between 272~ns to 328~ns. The power delay profile shows that LOS signal path is dominant for most cases since the distance between transmitter and receiver was relatively short less than 2~m. 


\subsection{Spectrum Division}
Low data rate application can be considered separately from high data rate applications, because the bandwidth and latency requirements are significantly different. Based on the extensive analysis of technical characteristics of WAIC, the spectrum requirements per aircraft are about $35$~MHz and $53$~MHz for both low data rate and high data rate applications, respectively~\cite{WAIC_spc}. Since WAIC system utilizes the $4.2-4.4$~GHz as the main spectrum~\cite{WRC15}, it is possible to divide the entire spectrum into two segments for  low and high data rate applications. In this way, we eliminate the co-existence issue between low and high data rate applications and simplify the design of entire WAIC systems~\cite{Gursu16}. For instance, WAIC may allocate $80$~MHz and $120$~MHz out of the total $200$~MHz band to low and high data rate applications based on the ratio of the spectrum requirements~\cite{WAIC_spc}. In fact, the 802.15.4 standard uses the $83.5$~MHz band by sub-dividing into 16 channels, where each one has a $2$~MHz bandwidth with a $5$~MHz channel separation in the $2.4$~GHz band~\cite{154}. By considering 802.15.4, we obtain around 15 channels using $80$~MHz for the low data rate applications. With a similar way, we have around 5-6 channels, each with a bandwidth of $20$~MHz over $120$~MHz for the high data rate applications. Hence, it is possible to resolve the co-existence issue between 802.15.4 and 802.11 frameworks by dividing the main spectrum. 

\subsection{Network Architecture}  
The WAIC network must have flexible and scalable architectures to support the requirements of various applications. Based on existing industrial wireless networks, modular and hierarchical networks make the system more flexible, robust, and reliable. 

\subsubsection{Network Element}
By considering industrial wireless standards of Section~\ref{sec:existing}, two main classes of WAIC devices can be defined: field devices and infrastructure devices. Each field device attached to the plant transmits the measurements to a gateway node. Then, the gateway eventually sends this measurement to the central onboard entity. Infrastructure devices may include gateway nodes, network manager, and central onboard entity, as shown in Fig.~\ref{fig:waic}. The network manager mainly configures the network and manages communication resources. The gateway node provides the bridge functionality, namely, the wireless connection on one hand to field devices, and the wired connection on the other hand to backbone networks. In addition, it can also collaborate with the network manager and manage the network resource for its field devices. To reduce the weight of WAIC equipment, the gateway node may be equipped with multiple communication transceivers. For instance, the Crossbow Startgate boards support both 802.15.4 and 802.11 standards for low and high data rate applications, respectively~\cite{crossbow}. While the WAIC framework does not explicitly define the specific type of backbone network, Internet Protocol (IP) networks based higher data rate avionics buses such as AFDX~\cite{A664} and Time-Triggered Ethernet~\cite{6802} are desirable to reduce the cost of the system integration.

\subsubsection{Topology}
The network topology of WAIC may consist of a number of sub-networks dependent on the transmit power and the RF environment, as shown in Fig.~\ref{fig:waic}. A gateway provides the network service of the sub-network, which can be extended using relay nodes. Generally speaking, the gateway only serves its compartment where it is located due to the significant possible signal attenuation between different compartments in the aircraft~\cite{WAIC_spc}. 

For instance, in Figs.~\ref{fig:waic} and~\ref{fig:RefAir}, a star topology with a gateway is the most appropriate for relatively small compartments such as the avionics and flight deck spaces. For the large compartments such as the cargo compartment or the passenger cabin, a multi-star topology may be considered to be more suitable in order to provide sufficient coverage by increasing the gateway nodes. Multiple gateway nodes provide several benefits such as reduction of traffic congestion and latency, and improvement of throughput and network reliability~\cite{whart}. 

In Fig.~\ref{fig:waic}, the star topologies can be also extended by using multi-hop line topologies without increasing the gateway nodes and transmit power for avionics, cargo, wings, and stabilizer compartments. In fact, the redundant routing paths of the multi-hop mesh topology improve  the higher level of fault tolerance compared to the ones using the typical star or the tree topologies for the large-scale industrial applications~\cite{Winter13}. The main advantage of mesh topologies is the absence of single-point of failures and easy re-configuration of the network to overcome temporary loss of communication links. However, a mesh network protocol generally brings serious challenges to implement and verify it due to its protocol complexity of individual nodes in practice~\cite{Park18}. Hence, a well-designed star topology may perform better than the mesh protocol for the enclosed compartment. WAIC must carefully investigate its benefits although the mesh topology is a preferable option in low data rate standards such as WirelessHART, ISA100.11a, and Bluetooth.

\subsection{Link Adaptive Scheme}  
WAIC systems require to adapt the modulation, coding, and other protocol parameters dependent on wireless channel conditions such as channel fading and interference. 

\subsubsection{Forward Error Correction}
Critical WAIC applications require extremely high reliable and low latency transmissions to meet the safety requirements~\cite{DO178B}. 802.15.4, Bluetooth, and 802.11 standards are generally configured to balance the tradeoff between throughput, latency, and energy efficiency. However, the typical target Packet Error Rate (PER) $0.01$ and $0.1$ of  both 802.15.4 and 802.11 standards, respectively, do not fulfill the flight safety requirements~\cite{Park17}. For instance, the possible target PER of critical actuating link is around $2 \times 10^{-4}$ based on the required target bit error rate of $10^{-6}$ for $224$ bit/packet in flight control systems~\cite{Elgezabal10}. 

The Forward Error Correction (FEC) scheme is one of the most suitable techniques to achieve high reliability and low latency~\cite{Holfeld16,Matsuzono10}. It is particularly desirable for time-critical sensing data and control signal transmissions of closed-loop controls since the general Automatic Repeat Request (ARQ) scheme does not meet both high reliability and low delay demands~\cite{Aikawa96,Liu97}. However, the overhead is still applied even if there is no error in the link~\cite{Choi06}. The default 802.15.4 standard does not include FEC while many 802.11 standards typically use turbo or Low Density Parity Check (LDPC) codes to improve the throughput, as shown in Table~\ref{tab:existing}.

Since many critical WAIC applications typically have a short data payload up to a few hundred bits,  the FEC options need to be optimized. Several channel codes, including convolution codes, Bose-Chaudhuri-Hocquenghem (BCH) codes, polar codes, turbo codes, and LDPC codes, are analyzed according to the block error rate, throughput, and complexity using practical decoders for ultra-reliable low latency communications~\cite{Shirvan19}. 

By considering FEC options of Table~\ref{tab:existing}, the convolutional code is one of the good candidates for real-time WAIC applications since it provides the comparable efficiency to LDPC and turbo codes for small block lengths. Due to error floor, turbo and LDPC codes are less efficient for very low value of the block error rate (e.g. $10^{-9}$), with respect to the ones of the convolutional code. Furthermore, in contrast to the typical iterative decoders of turbo and LDPC codes, the convolutional code provides a shorter decoding delay due to lower decoding complexity. The optimal decoding of BCH codes provides the most reliable performance due to the large minimum Hamming distance even under the low block error rate. However, they are not flexible since the block length and the information length can not be arbitrarily chosen.

\subsubsection{Modulation and Coding Scheme}
Modulation and Coding Scheme (MCS) is an interesting technique to achieve a high data rate. Higher modulation order and code rate increase the number of data bits per transmission while requiring more signal power because of risk of high error probability. IEEE 802.11n~\cite{11n} and IEEE 802.11ac~\cite{11ac} offer around 70 and 10 MCS options, respectively. 

Minstrel, the baseline MCS adaptation schemes for IEEE 802.11~\cite{11n}, can be used to optimize MCS selection of WAIC for non time-critical high data rate applications. However, the link adaptation algorithm may cause packet losses and jitter due to its suboptimal selections and MCS update variation. To guarantee the packet transfer, a fixed conservative MCS is recommended for time-critical applications. Furthermore, high signal-to-noise ratios demand with high order modulation is hard to achieve with low power design of energy-constrained WAIC nodes. Hence, the most robust and reliable MCS option is desirable such as Binary Phase Shift Keying (BPSK) with low code rate for the time-critical low data rate applications or energy-constrained transceivers.



\subsection{Hybrid Multiple Access}
To guarantee deterministic performance of avionics systems, FDMA and TDMA schemes are a natural option since they are inherently deterministic~\cite{Wang16, Park18}. A network manager can precisely assign a timeslot to transmit and receive a data based on the TDMA scheme of practical industrial standards in Section~\ref{sec:existing}. A timeslot is the smallest unit of time for communication. The latency becomes deterministic due to the reserved timeslot between nodes. In addition, it is possible to conserve energy consumption when the application has the low refresh rate. Each energy-constrained node turns its operations mode to the sleep mode when it does not evolve the communication. However, TDMA-based scheduling approach does not effectively handle the event-triggered traffic such as threshold-based pressure sensors. For such situations, a CSMA strategy might be more suitable. Furthermore, the CSMA scheme is effective for device joining, management, and retries. 

A hybrid combination of TDMA and CSMA is an attractive solution since it combines the advantages of both a deterministic access using schedule-based MAC and a random access using contention-based MAC~\cite{Rhee08,Pan11-thesis}. This hybrid protocol efficiently supports different requirements of heterogeneous applications. In fact, existing industrial wireless networks rely on the hybrid superframe structure combining both TDMA and CSMA access schemes~\cite{Park13_hybrid}. In each superframe, TDMA and CSMA schemes could be combined in several ways, one of which is higher-level TDMA slot scheduling and allowing for contention within the scheduled slots. While shorter superframes reduce the latency at the cost of more network bandwidth, longer superframes reduce the power and network bandwidth usage. When determining the superframe period, these tradeoffs need to be carefully considered~\cite{isa100}.


\subsection{Resource Allocation}
The resource allocation is a critical task to prevent unwanted interference from adjacent devices~\cite{whart,isa100}. 

\subsubsection{Slot Scheduling}
Network manager generates a timeslot schedule by considering the application requirements and the network states such as link quality and topology~\cite{Nhon15,Lu16}. Although WirelessHART and ISA100.11a networks rely on the centralized resource allocation for all network devices, the centralized approach can significantly degrade the adaptability due to the uncertainty of link conditions, large monitoring overhead, and reconfiguration cost. Note that industrial wireless standards do not describe how the resource allocation of the entire network should be implemented. Decentralized or distributed resource scheduling is considered to be more suitable for WAIC because of severe signal degradation between different compartments. The network manager may assign some portions of communication resources to the sub-network of the gateway. The gateway then allocates the timeslot and channel to its associated field devices. 

In real-time control loops, there is a time period for transmitting and receiving the data between controller and field devices to ensure the stability of control systems. Since the time-critical traffic needs to be delivered within a bounded latency, there is no need to send them after the time deadine has expired. In fact, the deadline based packet discard policy reduces the  computational and communication resources, and the memory consumption of the network. While IWLAN is designed to address time-critical high data rate applications, it possibly incurs the failure of real-time control due to the inefficient polling scheme~\cite{iwlan1}. In IWLAN, the length of all transmissions for one round becomes too long as increasing the number of nodes. By considering low data rate application, some scheduling algorithms or implementation details have been proposed for WirelessHART~\cite{Lu16,Sha17}, ISA100.11a~\cite{Palattella16,Park16_cloc}, and WIA-PA~\cite{Jin14,Jin18}. 

Since most high data rate monitoring applications do not need to operate continually at their respective maximum rates, it is possible to reduce instantaneous data rates by delaying non-critical sensing information. For instance, owing to the high sampling rate with high sample accuracy, the maximum data rate of each engine vibration sensor is expected to be $4.8$~Mbit/s~\cite{WAIC_spc}. Engine vibration sensors, however, have a very low duty cycle ($< $2\%), which means that the average data rate is around $80$~kbit/s. Measured data may be forwarded in the times between other high data rate applications such as imagery, in order to reduce the traffic variation.



\subsubsection{Frequency Hopping}
Multi-channel communication is an effective technique to improve the robustness against the multi-path fading and to reduce the interference impact from adjacent devices. The frequency hopping is included in the most promising wireless standards such as WirelessHART, ISA100.11a, and Bluetooth for low data rate applications. Simple frequency hopping is effective method to minimize interference and enhance security on congested channels for the outside applications. Recent low-cost wireless devices also support multi-channel communication~\cite{Wagner12}. However, the channel switching policy is still challenging since it must choose the least interfering channel while maintaining coordination to avoid the channel overlaps~\cite{Incel13}.







\subsection{Time Standard and Synchronization}
The network manager generally decides the common timeslot duration of the entire network in the industrial wireless networks~\cite{isa100,Chen10}. While shorter timeslots efficiently reduce the sampling interval for the short payload, longer timeslots obviously support longer payloads with possible extended functions including serial acknowledgements from multiple devices~\cite{isa100}. A flexible timeslot duration generally provides better network utilization based on the resource allocation optimization for heterogeneous applications. However, it can significantly increase the complexity of protocol stack implementation~\cite{Wang16}. The timeslot duration of WirelessHART is fixed to 10~ms~\cite{Chen10}, while its value in ISA100.11a is configurable such as 10~ms or 12~ms~\cite{isa100}. 

In the TDMA-based MAC operation, the fine-grained time synchronization is essential for devices to communicate with each other in the assigned timeslot. In fact, time-stamping an event based on time synchronization is useful in most control systems. Furthermore, slotted frequency hopping also requires tight time synchronization among neighbors. WirelessHART requires up to 10~ppm clock accuracy~\cite{Chen10}. However, clock skew between synchronized devices is inevitable because of voltage variations, temperature, or component aging even with highly stable crystals~\cite{iec_whart}. 

The WAIC system may combine centralized time synchronization scheme of WIA-PA and WLAN, and distributed time synchronization scheme of WirelessHART~\cite{whart}. In the centralized time synchronization mode of WIA-PA, the network manager is the original time source of the whole network. Hence, the network manager periodically transmit a specifically designed time synchronization frame to the gateways through the backbone network. Each gateway then synchronizes the time with its associated field devices by transmitting the synchronization frame. On the other hand, the pairwise time synchronization mode of WirelessHART propagates clock updates to their neighbors for the multi-hop mesh network. When communication is initiated between node pairs, time synchronization is performed between them. Keep-alive messages are still exchanged to maintain synchronization even if they do not have data to exchange.



\subsection{Multi-hop Routing} 
In Figs.~\ref{fig:waic} and~\ref{fig:RefAir}, the multi-hop mesh topology is particularly attractive for low data rate applications of the large-scale avionics, cargo, and wing compartments since an alternative path enhances reliability and robustness over various types of failures. Various routing protocols of WSNs have been proposed for traditional large-scale monitoring applications~\cite{Karaki04,Akkaya05,Watteyne11}. In practical industrial standards, all nodes of WirelessHART have the routing capability, whereas reduced function devices of ISA100.11a and WIA-PA standards do not support the routing capability. While the presence of reduced function devices degrades the robustness due to the limited number of alternative paths, there is a benefit in terms of  latency, energy efficiency, and complexity reduction of the protocol implementation. 

One of the most appropriate techniques is graph routing defined by both WirelessHART and ISA100.11a standards. Graph routing significantly enhances the worst-case reliability over typical source routing, whereas it degrades the latency and the energy efficiency~\cite{Mo15}. However, the experimental result shows that the typical latency of graph routing still ranges between $110$~ms and $2.8$~s even for a two-hop communication~\cite{Winter13}. Furthermore, it is still essential to efficiently integrate graph routing and resource scheduling algorithms to meet reliability and latency requirements for low data rate applications. 

Since the redundant packet transmission may cause the channel congestion and waste the network resources~\cite{Hardy08,Ishii09}, the flooding-based communication is typically avoided in industrial wireless standards. However, recent researches show that efficient flooding scheme substantially improves the reliability with minor overhead, if it is properly designed, over low-power and lossy networks~\cite{Ferrari11}. In fact, Bluetooth mesh uses a managed flooding scheme to take advantage of the inherent flooding redundancy while suppressing the excessive and unnecessary traffic for low data rate applications~\cite{PGWP}. The flooding scheme can be used for critical network configuration message or control signal for mission-critical applications.

In the 802.11 family, IEEE 802.11s standard provides multi-hop routing and forwarding mechanism for low power applications. As an ad-hoc infrastructure, this standard supports multi-hop communication between APs (up to 32 APs). Since the mesh network of 802.11s is implemented at the link layer, it relies on MAC addresses rather than IP addresses for its mechanisms. The default routing protocol called hybrid wireless mesh protocol uses a common set of protocol primitives based on ad hoc on-demand distance vector protocol~\cite{Perkins99}.

\subsection{Machine Learning-based Parameter Management}
\Blue{To support the time-critical operations, WAICs must carefully optimize a large number of parameters among different layers of the network. Mathematical input/output models have been typically used to select the set of parameters of communication systems. However, the interactions among the layers and practical limits such as interference and circuit nonlinearities are hard to model by simple expressions of adequate accuracy~\cite{Park13_hybrid}. Furthermore, the complexity of recent wireless systems can seriously increase as the result of MIMO, wide spectrum usage, natural disturbing factors, and various algorithms at different layers~\cite{11n,11ax,11ay}. The accurate prediction of reliability and delay is extremely challenging since these metrics depend on the wireless channel realization and the assigned network resource. Hence, the traditional adaptive resource management does not perform well in practical WAIC systems.}

\Blue{Deep Learning (DL) is an effective tool to solve several technical challenges of WAIC networks with large-scale topology and complex channel conditions. In various applications such as computer vision and natural language processing, DL has shown the impressive results on the pattern recognition from complex raw data~\cite{Goodfellow16}. Some recent papers provide an extensive survey of the possible applications of DL models at different layers of wireless networks~\cite{Mao18, Sun19, Zhang19_dl}.}


\Blue{One of the most promising tools is the online machine learning approach to select the communication parameters based on the real-time measurements of performance metrics such as reliability and delay~\cite{Sun19}. Online learning not only captures changes in the system model based on the data set but also learn the hidden features of system operation details. WAICs can provide several sets of network parameters where each set is associated with various performance metrics such as reliability and delay. Thus, adaptive resource management based on DL models is achieved by predicting the reliability and delay for each parameter set before the packet transmission~\cite{Carlo19}. Furthermore, reinforcement learning models are recently tied with many DL frameworks for resource allocation, routing, and network optimization since the communication system continuously interacts with the environment including the time-varying channel conditions~\cite{Aru17,Zhang19_dl}.}

\Blue{However, DL models are mostly used as black-box models without formal guarantees due to their nonlinear and large-scale nature. As a consequence, neural networks of DL models are vulnerable to input uncertainties or adversarial attacks~\cite{Moo17}. Such disruptions can be either of an adversarial nature~\cite{Szegedy14}, or merely compression and cropping~\cite{Zheng16}. These fundamental drawbacks limit the adoption of DL models in safety-critical applications~\cite{Boj16,Julian16,Xiang18}. Recently, many researches have been conducted for developing tools to measure or improve the robustness of DL models. Most works focus on specific adversarial attacks and attempt to harden the network such as crafting hard-to-classify examples~\cite{Goodfellow15, Alexey17, Moo16}. Although these methods are scalable and work well in practice, they still suffer from false negatives. Safety-critical WAIC systems require provable robustness against any bounded variations of the input signal. One of the most reasonable solutions is to use the data-driven DL approach as a complementary solution to traditional design techniques based on mathematical models of the wireless network design~\cite{Zappone19}. The mathematical models can significantly reduce the amount of actual data to implement data-driven DL approaches.}

\subsection{Effective Security}
Authentication and encryption of data using the AES-128 CCM cypher mode of operation is a de-facto standard for 802.15.4, Bluetooth and 802.11 networking solutions. This de-facto standard is sufficient to provide protection against brute-force attack for the typical traffic loads and time of operation (hours) for which the aircraft is vulnerable to attacks~\cite{security16}. 

On top of the networking data protection, it is recommended that each application domain may add 32-bit, 64-bit, or 128-bit message integrity check values. This provides different application security credentials for various security contexts whenever possible, so that even if some of the assets are compromised (i.e., because physically exposed), an attacker will gain access to only a small part of the whole WAIC system.

Refreshing the security credentials (i.e., by creating session keys) is necessary to prevent an attacker from being able to perform a successful brute-force attack on the security keys. For example, some standards recommends all credentials to be periodically refreshed once every 14 days~\cite{Bluetooth}. In WAIC, session keys can be valid for a single flight leg of an aircraft to improve the robustness. This is contrast to the widespread approach of the device deployment for which provisioning is done at the time of manufacturing, and devices are pre-configured with static security keys. At the same time, re-commissioning procedure and firmware updates need to be as simple as possible to avoid concealed code with risk of malicious actions. They also require thorough documentation and proper authorization.

\Blue{Concerning the network-level security, WAIC systems can benefit greatly from machine learning and DL in particular. Given a rather predictable and static flow of information available in WAIC, with a limited number of traffic classes (see Section~\ref{sec:waic}), the system can easily learn traffic patterns from measurements and generalize it to possible future anomalies, or identify patterns that deviate from normal behavior. This reduces the cost of pre-determined rules to distinguish intrusions~\cite{Zhang19_dl}. Traffic flow inference can be achieved at different layers, starting from the exploration of MAC-layer parameters (inter-packet arrival, packet size, etc), to higher layer information with more intrinsic features.}

\Blue{Moving further from the classification of security threats, DL is then greatly useful for intrusion detection in WAIC systems. For example, DL with auto-encoder offers 98\% accuracy in classifying four types of traffic in IEEE 802.11 networks, including flooding and impersonation traffic~\cite{Thing17}. In addition, DL with restricted Boltzmann machine has been proven effective with 99.1\% accuracy rate on intrusive behavior in a WSN-based network for safety-critical applications~\cite{Otoum19}.}

\begin{figure}[t]
  \centering
  \includegraphics[width = 0.95 \columnwidth]{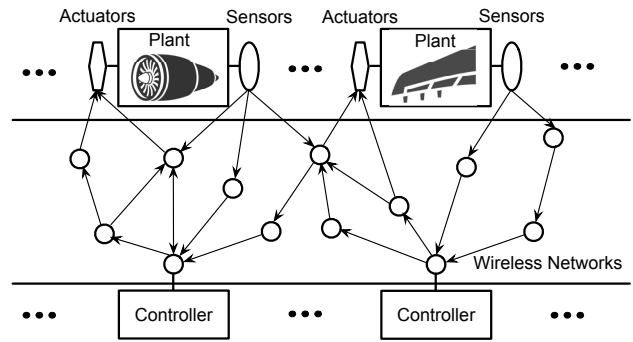} 
   \caption{\Blue{Overview of the conceptual WNCS setup based on WAICs. WAICs support the closed-loop control consisting of sensing links (from sensor to controller) and actuating links (from controller to actuator).}} \label{fig:wncs}
\end{figure}

\subsection{Joint Design of Control and Communication}
\Blue{A majority of current embedded wireless systems has largely focused on non-critical monitoring or open-loop control applications. When we apply the current common approach to critical closed-loop control applications, we face several technical challenges including tight timeliness and safety requirements, as discussed in Section~\ref{sec:cha}. We must re-think the communication architectures and protocols to maintain the control stability and performance even in the presence of disturbances to WAICs.}

\Blue{Wireless Networked Control Systems (WNCS) is a key approach for bridging the gap between control and communication aspects~\cite{Park18}. In Fig.~\ref{fig:wncs}, WNCS consists of spatially distributed sensors, actuators and controllers that communicate through WAICs, rather than traditional wired fieldbus connections. In most common approach of WNCSs, sensor nodes transmit the measurements of the plant state; controllers then compute control signal based on received plant state and send it to the actuators in order to affect the physical plant. Guaranteeing stability and safety of closed-loop control is an extremely challenging task mainly due to network uncertainties such as packet losses and delays.} 

\Blue{The control community typically solves the problem by approximating the details of the networks with tractable noise distribution and link loss models~\cite{Park18,Zhang01}. Although this mathematical-based approach guarantees stability and safety, it does not provide a systematic solution to the practical system design due to the unrealistic network assumptions. On the other hand, the network community mainly uses software to tackle the non-critical monitoring and open-loop control applications. However, it does not provide any guarantees on the stability and performance of critical closed-loop control~\cite{Park18}.}

\Blue{In practice, the dedicated controller typically imposes a set of requirements in terms of reliability, delay, or update deadline of the network~\cite{Park18, Park16_cloc, Park11}. However, since the assignment of network resources and routes is a generally static setup, it requires global reorganization with large overheads to adapt dependent on packet losses, delays, and faults. In addition, the control configurations must be re-computed dependent on the network changes~\cite{Park18, Alur11}. Moreover, physical node-level programming of WNCSs is one of the fundamental reasons for poor robustness of the overall control systems~\cite{Welsh04}.}

\Blue{The possible research direction is to apply the distributed computation paradigm to the WNCS design based on WAICs. The entire network itself serves as a controller by spreading the computation of the control algorithm, instead of assigning a specific node as the controller node. In other words, the control functionality can be decoupled from the physical node. In the presence of unexpected network changes, it could efficiently migrate the control tasks to the most reliable set of candidates for maintaining the stability and safety of aircraft. Furthermore, even if a single sensor is unavailable due to link losses, faults, or attacks, the nodes should cooperate to compute the control signal without interruptions.}

\subsection{Fault-Detection and Fault-Tolerant Design}
\Blue{To meet strict safety demands, WAIC must provide efficient fault-detection, isolation, and recovery mechanisms~\cite{Yeh01, Buckwalter08}.  However, the major avionics systems still have the manual and template-driven scheme for the fault-detection, isolation, and recovery~\cite{Yeh01, Lee90}. While the DL method could considerably improve the fault-detection and diagnosis performance, the inference and training still require significant computational resources to run in practice~\cite{Goodfellow16}. Recent integration of DL and edge computing essentially solves the problem of the high-computation and low-latency requirements of the data-driven fault-detection and diagnosis~\cite{Chen19}. Edge computing is a distributed computing scheme in which information processing based on DL models is locally performed to improve the response time, bandwidth efficiency, and scalability. However, distributing the logic into various network nodes introduces new challenges~\cite{Wang20}.}



Avionic systems generally have multiple redundancies in terms of sensing, communication, and computation to continue performing its intended operations in spite of subsystem faults~\cite{Buckwalter08}. In WAIC, each redundant can be connected using a wired or wireless setup to its primary device as a backup. The redundant device must continuously provide an acceptable level of services by taking over the function of the primary device, if  the primary one is not functional. Hence, WAICs need to reconfigure the network to take over the data transfer with secondary systems during primary system failure. Fortunately, avionic system typically reduces the level of service demands once a primary system malfunction has been detected since the performance requirements of primary systems are too strict~\cite{Lee90}. 

In WAIC, field devices must be distributed over different space around a set of plants while maintaining the robust connectivity, in order to avoid a complete control failure. Hence, the node placement is a complex problem as a function of the locations of control process, physical dimension, and network topology. Given a network topology graph, the degree of connectivity is the most commonly used robustness metric to quantify the network resistance to random failures or attacks~\cite{Shi18,Hu18}. WAIC nodes should be deployed systematically to improve the network fault tolerance, whereas many research papers of WSNs assume a randomly deployed nodes.

\subsection{\Blue{Prediction and Adaptation for Natural Disturbances}}
\Blue{The most relevant natural disturbances of WAICs are atmospheric gases, hydrometeors, electrometeors and solar activities~\cite{Aglargoz14}. Some ITU reports provide the signal attenuation model of various natural phenomena~\cite{alt_1,alt_2,rain_1,fog}. The model of~\cite{alt_1,alt_2} calculates the signal attenuation as a function of atmospheric gases for different altitudes. This signal attenuation model basically includes the pressure, temperature, and water vapor properties of a standard atmosphere.  Besides, it is also possible to integrate varying signal degradation due to rain rate with altitude~\cite{rain_1, rain_2}. Furthermore, the effect of the cloud and fog on the signal attenuation is modeled based on the statistical data of total columnar content~\cite{fog}.}

\Blue{We implement the signal attenuation model of the standard atmosphere~\cite{alt_1,alt_2}. The model does not include the path loss effect to avoid the variations depending on the location of the node. The upper altitude boundary is set to 12,000~m based on the maximum operation altitude of A320~\cite{alt_3}.}

\begin{figure}[t]
  \centering
  \psfrag{x}[][]{\footnotesize{Frequency (GHz)}}
  \psfrag{y}[][]{\footnotesize{Altitude (m)}}  
  \psfrag{z}[][]{\footnotesize{Signal attenuation (dB)}}    
  \includegraphics[width = 0.95\columnwidth]{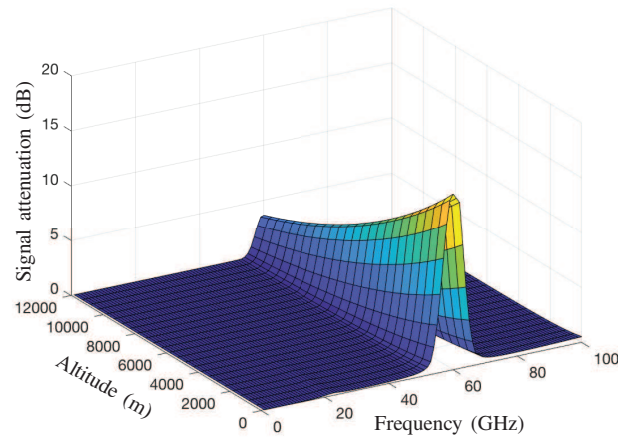}
   \caption{Signal attenuation as a function of different operating frequencies $1-100$~GHz and altitudes $0-12,000$~m.} \label{fig:alt}
\end{figure}


\Blue{Fig.~\ref{fig:alt} shows the signal attenuation as a function of different transmission frequencies $1-100$~GHz and altitudes $0-12,000$~m. The main peak is around $57-60$~GHz due to resonance of Oxygen~\cite{alt_2}. In fact, the received power is reduced by roughly $14$~dB. The second peak is observed around 22~GHz, which corresponds to the resonance of water vapor. It is significantly lower than the main peak. Even though the attenuation gets higher as decreasing altitude due to the atmospheric gases, its effect below 10~GHz is negligible.}




\Blue{While the signal degradation due to atmospheric gases and hydrometeors is shown to be negligible on the primary spectrum $4.2-4.4$~GHz of WAICs, both natural disturbances significantly degrade the signal propagation in the 60~GHz millimeter waveband of spectrum~\cite{Aglargoz14}. Hence, some new standards such as 802.11ad and 802.11ay must carefully adapt their operations dependent on altitudes and weather conditions.}

\Blue{Although the electrometeors and solar activities obviously affect the avionics systems, the effect analysis on the signal attenuation is not yet available. Furthermore, these natural disturbances are extremely rare events compared to the entire operation time of flights. Some recent researches apply the machine learning techniques with a physical understanding of the environment to predict the electrometeors and solar activities~\cite{Inc18, McGovern17}. WAIC must support the time-critical operations to guarantee the safety and stability of aircraft even without a channel model.}

\Blue{End-to-end learning of communications systems is an interesting novel concept for learning full implementations of transmitter and receiver by considering a specific performance metric and channel model~\cite{Shea17}. The basic idea is to treat the entire communication system as an autoencoder and to model transmitter and receiver as neural networks~\cite{Goodfellow16}. Then, the entire model is trained using stochastic gradient descent as the supervised learning framework. However, end-to-end learning still requires information about the gradient of the channel transfer function to train the DL models. One possible research direction is to explore the field of deep reinforcement learning to overcome the missing gradient of any channels without prior knowledge.}

\subsection{Effective Directional Antenna and MIMO}

\subsubsection{Directional Antenna}
In a harsh WAIC environment, the blocking or multi-path limits the useful amount of radio signals. Directional antenna design is desirable to enhance network performance while reducing power consumption at certain WAIC nodes. Higher gain antennas reduce the delay spread and amplify the primary transmission path. Furthermore, it decreases the interference effect to other aircraft unless they are aligned for outside applications. In addition, energy-constrained nodes can further reduce its transmit power by using higher antenna gain. However, a large portion of the transceivers have restrictions on the physical size due to weight limitations and installation locations. Hence, WAIC systems may consist of heterogeneous classes of gain antennas.

\subsubsection{MIMO}  
Antenna diversity is another suitable technique to enhance the reliability and throughout over a fading channel. MIMO systems rely on multiple vector antennas of both the transmitter and the receiver~\cite{Zheng15}. While the spatial multiplexing considerably improve the throughput of IEEE 802.11n/ac, the spatial diversity technique is desirable for mission-critical WAIC applications. In the MIMO diversity technique, the same data is transmitted across the different antennas to improve the reliability since independent fading paths have the low correlation against deep fading. For instance, to meet the low outage probability of $10^{-9}$, the required margin 90~dB is reduced to 18~dB and 9~dB by using diversity orders 8 and 16, respectively~\cite{Johansson15}. 

In an OFDM system, the frequency and spatial diversities are the powerful tool to spread the coded bits for achieving high reliability. If the Hamming distance between correct and erroneous code is $d$, these $d$ symbols must be assigned to different frequency or antennas. In~\cite{Park19}, we combine the STBC and the maximum ratio combining at the multi-antenna transmitter and the multi-antenna receiver, respectively. The significant improvement of link reliability is obtained by combing MIMO and STBC based on experimental channel measurements of aircraft.

\subsection{Efficient Integration}  
While WAIC needs to be connected to central control systems through high data rate buses, it does not explicitly define the specific type of backbone networks~\cite{WAIC_obj}. As avionics systems progress, there is a strong desire to move to more Ethernet-based networks in order to support the high data rate demand while reducing the development cost~\cite{Alena07,Buckwalter08}. Hence, Ethernet-based fieldbus network such as AFDX~\cite{A664} and Time-Triggered Ethernet~\cite{6802} are good candidates since many industrial wireless standards are also IP-compatible. 

Since there are major differences between wireless networks and these avionics data buses at both the MAC and PHY layers, a gateway must integrate the data bus network interface, acting as a wireless network AP~\cite{Cena08}. 

\subsubsection{802.15.4, Bluetooth and 802.11 Adaptation}
6LoWPAN and 6LoBTLE provide an IP interface for 802.15.4 and Bluetooth, respectively~\cite{Hui10}. The adaptation layer allows the direct communication of 802.15.4 and Bluetooth devices to other IP devices. 6LoWPAN and 6LoBTLE include two main functions, namely, the packet size adaptation and the address translation between these networks~\cite{Higuera11}. 802.15.4 (resp. Bluetooth) nodes still have a short 16-bit (resp. 48-bit) address instead of a long IPv6 address (128-bit) to simplify the operation and save the bandwidth. In WAIC, the gateway could convert the address between a short address of low data rate networks and a 128-bit address of the IP-based backbone network.

IEEE 802.11 uses unique 48-bit address as the IEEE 802 standard family. Since both 802.3 Ethernet and 802.11 WLAN use the same logical link control as defined by 802.2, the 802.11 and Ethernet-based fieldbus networks are relatively simple to integrate. 

\subsubsection{Wireless Backbone Networks}
It is possible to replace fieldbus backbone to emerging high data rate WLAN standards. The wireless backbone is particularly attractive to further reduce the weight of small unmanned aerial vehicles since the weight contribution of the backbone is significantly greater than that of sensing and actuating connections~\cite{Gud14_struct}. Furthermore, the safety requirement of small unmanned aircraft is relatively lower than the one of manned aircraft~\cite{Loh09}. 

\Blue{IEEE 802.11ad~\cite{11ad_mm} and 802.11ay~\cite{11ay} support the WLAN communications in the $60$~GHz spectrum. Since the maximum transmission rate of 802.11ay (resp. 802.11ad) is 40~Gbit/s (resp. 7~Gbit/s), this technology is a good candidate  for the backbone WAIC network. Note that 802.11ay archives much higher transmit rates than the one of 802.11ax due to the larger bandwidth. However, both 802.11ad and 802.11ay standards must carefully adapt their parameters depending on the atmospheric gases and hydrometeors. Furthermore, these standards are hard to penetrate any types of walls in the 60~GHz spectrum~\cite{Carlo16}.}

\subsection{Power Delivery and Data Transmission Through Metals}
The conducting nature of the aircraft prevents the RF signal propagation because of metal skin effect~\cite{WAIC_spc, Park19}. To maintain structural integrity, it is often not desirable to drill holes between bulkheads of aircraft when we install the WAIC nodes. Furthermore, the periodic battery replacement is impractical for energy-constrained devices within sealed containers of aircraft. 

Piezoelectrics, inductive coupling, and electromagnetic acoustic transducer is an appropriate technique to communicate and deliver power through a metal wall~\cite{Yang15}. Using inductive coupling, a power transfer efficiency is around 4\% through 20-mm-thick stainless steel. The electromagnetic acoustic transducer provides the $1$~Mbit/s data rate for through-metal communications~\cite{Graham11}. In~\cite{Wanuga12}, the OFDM ultrasonic PHY layer achieves the data rate of up to $15$~Mbit/s through metal walls. Furthermore, the underlying physics of sheet may be used to produce surface waves at metal-air interfaces for power and data transmission~\cite{Oruganti16}. 
\section{Conclusions}\label{sec:con}
Wireless avionics intra-communications bring significant advantages to improve efficiency and flexibility while reducing the costs of installation and maintenance over traditional fieldbus networks of aircraft. In this article, we have surveyed opportunities, technical challenges, and fundamental design principles to develop and deploy the emerging wireless avionics intra-communications for aircraft. We described the safety level of various critical operations and the current state-of-art of the frameworks in terms of system classifications based on data rate demands and transceiver installation locations. We then discussed the major technical challenges for realizing the envisioned aircraft applications and raised the issues related to the design of the communication protocols. The wireless avionics intra-communication basically needs to be cost-efficient while providing comparable real-time performance to current fieldbuses over harsh and complex aircraft environments. Furthermore, existing industrial wireless standards were briefly discussed to investigate the feasibility of current solutions to support the critical heterogeneous applications. Among the candidate technologies, 802.15.4 and Bluetooth have been considered for the low data rate applications, 802.11 for high data rate applications. Based on the candidate standards, we suggested fundamental schemes and design choices in terms of network architecture, protocol, and resource management to support the critical aircraft operations, indicating the most promising research directions in this field.


\bibliographystyle{IEEEtran}
\bibliography{ref}

\end{document}